\documentclass[preprint,12pt]{elsarticle}

\usepackage{amssymb}

\usepackage{siunitx}
\usepackage{subcaption}

\usepackage{setspace}

\journal{Technological Forecasting \& Social Change}

\begin{document}

\begin{frontmatter}



\title{Modeling the Evolutionary Trends in Corporate ESG Reporting: A Study based on Knowledge Management Model}


\author[label1]{Ziyuan Xia}

\affiliation[label1]{organization={Antai College of Economics and Management, Shanghai Jiao Tong University},
	            addressline={1954 Huashan Road}, 
	            city={Shanghai},
	            postcode={200030}, 
	            state={Shanghai},
	            country={China}}

\author[label2]{Anchen Sun}
\author[label2]{Xiaodong Cai}

\affiliation[label2]{organization={Department of Electrical and Computer Engineering, University of Miami},
	 	addressline={1251 Memorial Drive}, 
	 	city={Coral Gables},
	 	postcode={33146}, 
	 	state={FL},
	 	country={USA}}
	 	
\author[label1]{Saixing Zeng}

\begin{abstract}
Environmental, social, and governance (ESG) reports are globally recognized as a keystone in sustainable enterprise development. However, current literature has not concluded the development of topics and trends in ESG contexts in the twenty-first century. Therefore, We selected 1114 ESG reports from firms in the technology industry to analyze the evolutionary trends of ESG topics by text mining. We discovered the homogenization effect towards low environmental, medium governance, and high social features in the evolution. We also designed a strategic framework to look closer into the dynamic changes of firms’ within-industry scores and across-domain importances. We found that companies are gradually converging towards the third quadrant, which indicates that firms contribute less to industrial outstanding and professional distinctiveness in ESG reporting. Firms choose to imitate ESG reports from each other to mitigate uncertainty and enhance behavioral legitimacy.

\end{abstract}


\end{frontmatter}


\section{Introduction}
ESG (environmental, social, and governance) reports are essential to corporate business strategies since they provide a comprehensive assessment of a company's sustainable development, such as environmental management, social issue practice, and governance structure~\cite{lokuwaduge2017integrating}. ESG reports inform stakeholders about the conformity of corporate activities with ethical standards, assisting their understanding of equity value creation, risk management, and firm performance~\cite{giese2019foundations, HUARNG2024114327}.
ESG reporting is essential for sustainable investment, providing transparency, accountability, and comparability by integrating environmental, social, and governance factors into investment decisions~\cite{bouslah2023csr, tao2023corporate}. It enables investors to assess ESG performance, identify risks and opportunities, and benchmark companies across sectors and regions~\cite{tamimi2017transparency, giamporcaro2020orchestrating}. ESG reports are important for companies looking to attract investment capital from institutional investors, such as pension funds, which prioritize ESG considerations~\cite{barros2024esg}. ESG reporting is crucial for enhancing transparency, fostering responsible investment practices, and steering companies towards sustainable development~\cite{beccarini2023contingent}.

The three  elements of environmental, social, and governance play a critical role in determining the direction and performance of organizations in the area of corporate strategy~\cite{LI2023113648}. The environmental dimension of a corporation covers key topics, such as carbon emissions, climate change, and waste management, detailing the ecological impact of the firm~\cite{guoyou2013stakeholders}. By strategically addressing environmental concerns, businesses not only contribute to sustainability but also access cost-saving opportunities through efficient resource utilization~\cite{alsayegh2020corporate}. The social dimension examines how a corporation practices social responsibilities, covering topics such as data security, wellness and well-being, and responsible sourcing~\cite{kim2023corporate}. This factor has a significant impact on firm reputation, staff satisfaction, and customer loyalty, supporting long-term sustainability and market competitiveness~\cite{liu2023does}. Finally, the governance dimension evaluates the firm's internal governance, organizational behavior, and ethical oversight at the institutional level~\cite{oh2022trees}. Effective governance promotes investors' confidence and risk reduction by encouraging openness, accessibility, and smart decision-making~\cite{MA20171365}. Businesses may improve their resilience, reputation, and stakeholder value while putting themselves at the forefront of ethical business practices in an environment that is becoming scrutinized to a greater extent, by aligning their strategic goals with these ESG characteristics~\cite{ferrell2021addressing, bae2022can}.

ESG reporting holds crucial importance for technology firms due to its role in aligning business operations with evolving societal expectations, regulatory landscapes, and stakeholder demands~\cite{IURKOV2024114395}. Given the rapid pace of change in the technology industry, where new trends and challenges regularly emerge, ESG reports provide a framework for tech firms to navigate these dynamic shifts while demonstrating responsible conduct, managing risks, and maintaining resilience in a highly competitive and transformative environment~\cite{alkaraan2022corporate}. Technology companies should prioritize ESG issues in order to proactively manage the possible demands of governmental legislation and public opinion, protecting their reputational capital, regulatory compliance, and sustainable growth within the intricate and closely watched environment of the technology industry~\cite{egorova2022impact}. To strategically capitalize on opportunities for increased publicity and the cultivation of a positive market image, including initiatives such as ESG investment and the integration of ESG principles into their business strategies, technology firms are required to monitor and engage with emerging ESG topics~\cite{goodman2015engaging, freiberg2020esg}. By doing so, they build stakeholder trust and position themselves as responsible and forward-thinking corporate entities within the business environment.



The myriad of diverse reports available in the market poses a significant challenge for management researchers~\cite{sundararajan2013research}. The demand for extensive human involvement in data collection, identification, measurement, knowledge transfer, and analysis necessitates substantial resources for management research~\cite{XIE20165210}. Particularly in the 21st century, following the advent of the internet, we are witnessing an era of data explosion~\cite{chen2012business}. Consequently, there has been an increasing shift toward utilizing computational science technologies to create frameworks capable of automated data analysis~\cite{shaw2001knowledge}. Furthermore, advancements in natural language processing have expanded our capabilities in regard to text-based information extraction~\cite{xia2021mining}. The current study has been designed and conducted in response to this rapid development and to further progress ESG research.

\begin{figure}
	\centering
	\includegraphics[width=0.8\textwidth]{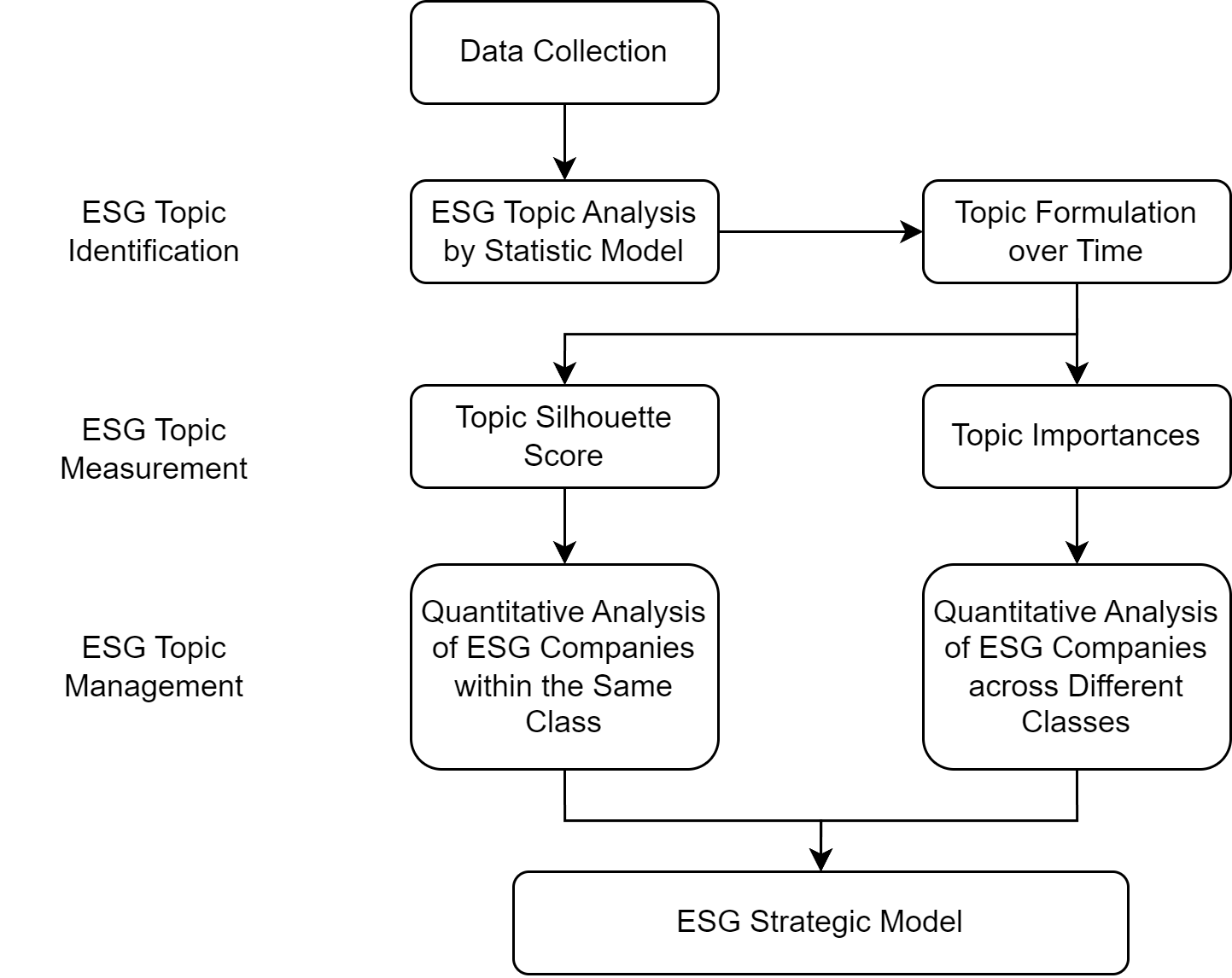}
	\caption{The Proposed Fundamental ESG Strategic Framework Diagram}
	\label{fig:overalldiagram}
\end{figure}

This study sets out with a primary objective to develop an effective approach to dissecting and understanding the evolution of topics found within technology firms' ESG reports. The overarching goal is to unearth latent dynamic patterns, providing actionable insights that can guide informed managerial decisions. We initiate this process by designing a data preparation module, responsible for crawling ESG reports from open-source networks. These reports are then pre-processed using natural language processing technologies in order to construct our comprehensive ESG dataset. We further delve into the dataset through a multi-faceted analysis, aiming to garner a spatial understanding that translates into quantitative managerial guidance. The resultant quantitative findings subsequently inform the construction of a detailed ESG strategic model. This model acts as a vital tool for visualizing each company's position in regard to ESG development and guiding the formulation of bespoke strategies. This research introduces a pioneering framework, as visualized in Figure.~\ref{fig:overalldiagram}, devised to analyze ESG reports and develop appropriate strategies tailored to diverse business scenarios. We illuminate the evolving trends of topics by extracting crucial text content from ESG reports of technology sector companies.

Our study proficiently identifies and analyzes the evolving patterns inherent in ESG topics. This research enriches institutional isomorphism theory, elucidating firms' inclination towards the pursuit of conformity rather than differentiation, thus giving rise to a discernible trend of clustering within a shared quadrant~\cite{dimaggio1983iron}. Within the realm of technology corporations, we unveil the phenomenon of ESG homogenization, wherein a significant portion of these entities' reporting practices progressively align with established norms, a tendency that potentially compromises their capacity for innovation and distinctive engagement.

The structure of this paper unfolds as follows: Section~\ref{sec:litr} opens with an introduction to contemporary ESG management research, emphasizing the integration of machine learning and other computational technologies in the realm of business strategies. Subsequently, in Section~\ref{sec:framework}, we elucidate our innovative framework and elaborate on the step-by-step procedure for the creation of the ESG strategic model. In Section~\ref{sec:exp}, a comprehensive compilation of 21st-century ESG reports from technology firms is deployed to substantiate the effectiveness of the proposed framework. The concluding remarks are presented in Section~\ref{sec:con}, where potential avenues for future research building on this study are also discussed.

\section{Literature Review}\label{sec:litr}
\subsection{ESG Reports and ESG Management}
ESG reports have evolved from environmental or sustainability reports since the turn of the century. In the first decade of this century, researchers and companies' perceptions of sustainability have centered on the dimension of environmental protection~\cite{arvidsson2022corporate}. However, with the continuous development of information technology and the emergence of new technologies and business models, our understanding of sustainability in both the industry and academia is also evolving~\cite{mervelskemper2017enhancing}.In recent years, ESG reports have become increasingly diverse in terms of topics and perspectives~\cite{fatemi2018esg}.

At present, investors and institutions are increasingly acknowledging the criticality of environmental, social, and governance dimensions in appraising prospective investments\cite{BAKER2021101913}. Literature indicates the positive correlation between ESG incorporation and a firm's financial and stock market performance~\cite{https://doi.org/10.1002/bse.3500}. ESG management not only drives a company's financial performance and mitigate risk, it is also a catalyst for the alignment of investments with sustainable development goals, resulting in positive environmental and social outcomes~\cite{egorova2022impact, https://doi.org/10.1002/bse.3500}.

In the sustainable management research landscape, ESG reports are perceived as instrumental tools that facilitate a comprehensive understanding of the complex interplay between business practices and ethical goals~\cite{terjesen2016board, tan2021ethics}. ESG reports provide meticulous documentation of corporate sustainability across all dimensions, thus providing a more comprehensive framework for scholars to assess corporate sustainability performance. ESG reports thus provide researchers in management with a large database and diverse sources enabling exhaustive analysis~\cite{english2023can}. With this capability, scholars can examine the impact of various ESG topics on corporate financial performance, green performance, and innovation capabilities~\cite{zyglidopoulos2016corporate}. However, the existing literature lacks a dynamic analysis of the evolution of overall ESG trends. Our study will fill this gap, thus providing a more intuitive and quantitative pattern of ESG evolution for management science.

As a result of organizational learning processes, growing stakeholder awareness, and introduced regulatory pressure, it is anticipated that ESG reporting will evolve significantly~\cite{alkaraan2022corporate}. Organizational learning enables businesses to modify and enhance their sustainability strategies based on emerging perspectives and prior experiences, resulting in more advanced ESG reporting~\cite{ZHANG2023122114}. At the same time, stakeholders' expectations for accountability and transparency in reporting rise as they become more conscious of environmental, social, and governance concernss~\cite{asif2023esg}. Moreover, ESG reports are significantly shaped by regulatory requirements. Companies need to comply with increasingly stringent standards pertaining to sustainability reporting from governments and international organizations by providing more precise and comprehensive ESG data~\cite{10.1093/rfs/hhz137}.


\subsection{Text Mining and Analysis}
The dawn of the data era, characterized by a rapid accumulation of data resulting from continuous advancements in data storage technology and internet infrastructure, is predominantly witnessed among leading and listed companies~\cite{goh2013social}. These entities are generating a wealth of textual data, encapsulating their operational environment, strategic developments, and social responsibilities~\cite{melville2010information}. This unprecedented data explosion introduces new prospects in regard to investigating the sustainability of corporations and the industry in general~\cite{watson2010information}.

Historically, research methodologies in business management engineering have played a critical role in understanding an array of corporate strategies and developments~\cite{mithas2011information}. However, these traditional approaches are inadequate when substantial text content analysis is invovled. To bridge this gap, text mining, through the utilization of machine learning, natural language processing, and other digital technologies, offers promising capabilities~\cite{greenwald2010evaluating,meyer2014machine,wang2020mining}. By extracting patterns, laws, and trends from voluminous textual data, text mining facilitates a quantitative approach to management engineering research, signaling a significant paradigm shift in the field.

Within this realm, the Term Frequency-Inverse Document Frequency (TF-IDF) is a commonly applied weighting technique that is widely recognized in information retrieval and data mining practices~\cite{ramos2003using}. The essence of the TF-IDF methodology lies in its ability to identify words or phrases that appear frequently within a single document but are scarce elsewhere, signaling a meaningful categorical difference and thereby serving as an effective classification tool.

In the existing body of literature, research by Lee et al. (2023) proposes an ESG information classifier to address the current limitations of ESG ratings, verifying the model's effectiveness through several application experiments and suggesting future work on integrating non-financial performance into ESG ratings~\cite{lee2023esg}.

In addition to classification tasks, Zanatto et al. (2023) analyzes the influence of ESG news on the volatility of the Portuguese stock market, establishing a custom News Sentiment Index to capture positive and negative ESG news related to companies listed on the PSI-20. Their findings indicate that ESG news helps reduce volatility during non-crisis periods but fails to have a significant effect during economic downturns~\cite{zanatto2023impact}.

Simultaneously, Cheng et al. (2023) investigated the correlation between brand reputation, CSR performance as measured by ESG scores, and the issuance of green bonds. The study conclude that superior ESG scores and brand reputation jointly drive the successful issuance of green bonds~\cite{cheng2023interactive}.

Finally, Lu et al. (2023) examine the relationship between digital transformation and a firm's ESG performance using Chinese-listed firms as a case study~\cite{ludigitalization}. Their research suggest that digital transformation fosters a firm's ESG performance by improving internal control and promoting green innovation, particularly for non-state-owned enterprises, manufacturing industry firms, high-tech companies, and companies with higher degrees of independent directors and analyst coverage. This research enriches our understanding of the underlying mechanisms linking digital transformation and ESG performance.

\section{Proposed Framework}\label{sec:framework}

This section delineates our innovative framework, meticulously crafted to streamline the processing and analysis of ESG reports from a wide range of companies across multiple industries. This framework converts these reports into quantified, actionable ESG thematic data, enabling researchers to perform comprehensive quantitative analyses of sustainability development strategies prevalent in diverse sectors. Further, our framework aggregates ESG topics over time, and utilizes clustering models that mining within-industry representiveness and classifier model that incorporate cross-sector distinctiveness. This approach facilitates the tracking of evolutionary trends within our strategic model, significantly broadening the scope and depth of ESG research. Ultimately, this framework serves as a crucial resource for researchers committed to advancing sustainable practices across various industries.

\subsection{ESG Data Collection}

\begin{figure}[h]
	\centering
	\includegraphics[width=0.7\textwidth]{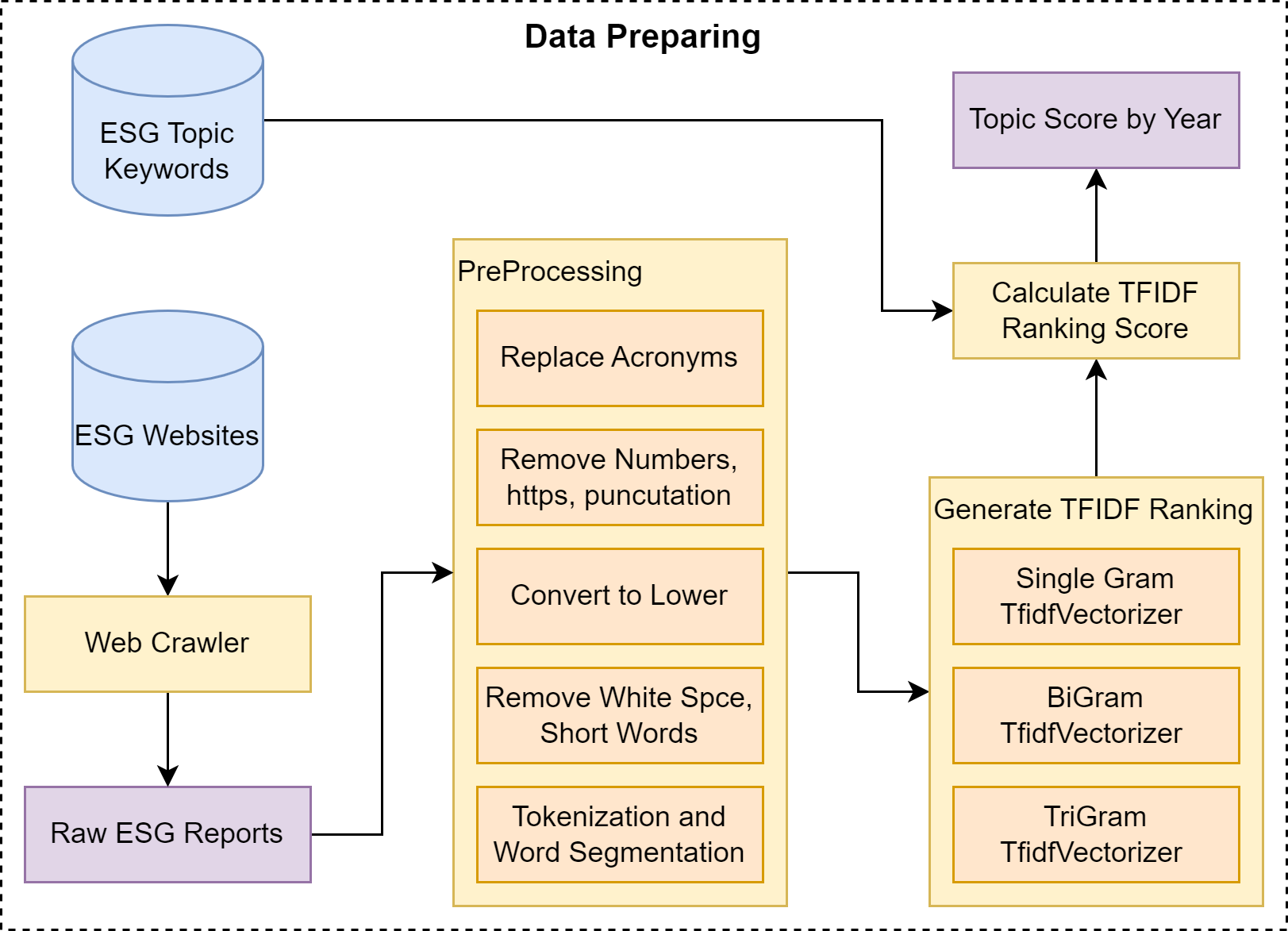}
	\caption{The Proposed ESG Data Preparing Framework}
	\label{fig:esg}
\end{figure}

In our current comprehension of the ESG landscape, the lack of availability of an open-source ESG thematic database constitutes a formidable barrier for researchers. This limitation hinders comprehensive examinations of the industry and primary areas of concern pertaining to ESG sustainability development over time and at the corporate level. To bridge this gap, we propose an ESG thematic data processing framework, illustrated in Figure.~\ref{fig:esg}, which draws upon ESG reports from a diverse range of companies.

Our groundbreaking framework enables the seamless processing and analysis of ESG reports sourced from numerous companies across various industries. It consequently generates quantified, actionable ESG thematic data. Researchers can harness this data to conduct quantitative analyses of sustainability development strategies prevalent across distinct industries, as discussed in greater detail in Section~\ref{sec:idenR}. This innovative proposition marks a substantial advancement, enhancing the depth and scope of ESG research, and providing researchers with an invaluable tool. 

\subsection{Identification of ESG Topics}

In our approach to extracting and analyzing ESG data, we systematically gather ESG reports from numerous companies over various years, forming a detailed data structure that captures the year, the company, and the specific ESG topics addressed. To accurately determine the significance of each ESG report, we employ the Term Frequency-Inverse Document Frequency (TF-IDF) algorithm. Unlike simple keyword counts or weighted keyword appearances that might inflate the importance of frequently used but less informative words, TF-IDF provides a more refined measure that evaluates how crucial a word is to a document within a larger corpus. This method enhances the discrimination of key ESG topics by diminishing the weight of common terms across all documents and amplifying the significance of words that are crucial in specific reports.

Our framework incorporates several text preprocessing techniques to ensure high-quality data analysis. We begin by replacing acronyms with their full names, as defined in our supplementary ESG acronyms dictionary, to maintain consistency. We then remove numbers, URLs, punctuation marks, excessive whitespace, and short words, while converting all characters to lowercase. Following these initial steps, we conduct tokenization and word segmentation to ensure the extracted text from the ESG reports is clean and precise.

Ultimately, this structured approach to data preprocessing and the application of the TF-IDF algorithm enable us to produce ranked ESG scores that are both accurate and meaningful. This methodology allows for a nuanced tracking of ESG topic evolution over time and facilitates the generation of various analytical plots (see Figures~\ref{fig:ESGplots} and~\ref{fig:Topicplots}). This improved framework not only aids in a more precise understanding and comparison of ESG topics within and between industries but also supports effective management and policy-making decisions by highlighting relevant and significant ESG trends.

\subsection{Measurement of ESG Topics}\label{sec:meas}

Our innovative framework delineates two distinct dimensions for quantifying ESG topics. Initially, we focus on analyzing ESG strategies within the context of industry representativeness. Following this, we examine the strategies in terms of their cross-sector distinctiveness. Each of these analytical trajectories is thoroughly explored, with detailed procedural flows for both dimensions described in the subsequent sections.

\subsubsection{Within-Industry Representiveness (x-Axis)}
\begin{figure}[h]
	\centering
	\includegraphics[width=0.6\textwidth]{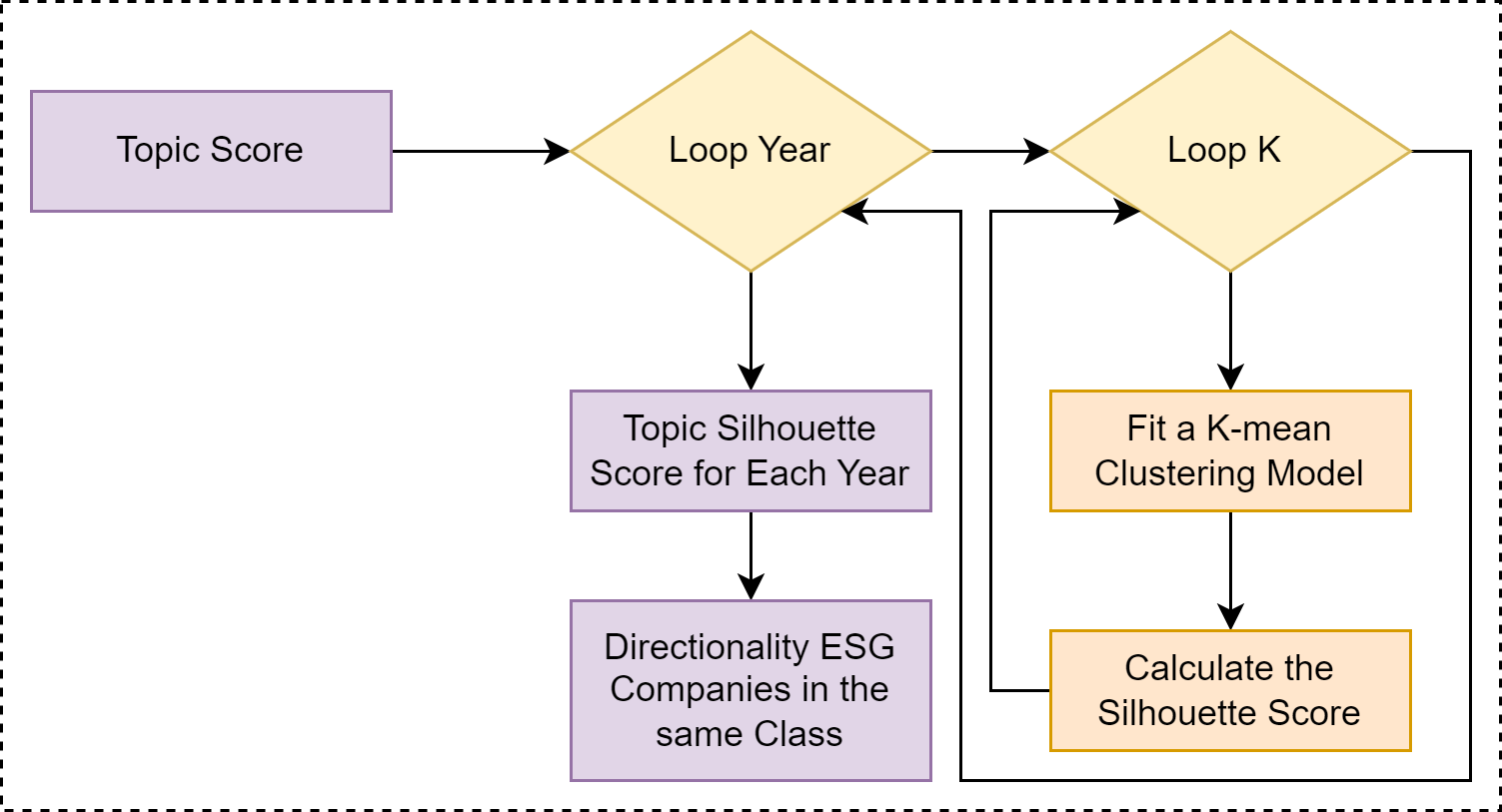}
	\caption{Scrutiny of Directionality for ESG Companies Within-Industry Representiveness}
	\label{fig:sameind}
\end{figure}

Our framework enhances the evaluation of ESG topics for companies within the same industry by effectively leveraging ESG topic scores previously identified. The methodological pipeline is visually represented in Figure~\ref{fig:sameind}. It begins with the parsing of each year's ESG topic data, followed by the application of a K-means clustering model. In this model, companies are treated as individual samples, and each ESG topic is considered a distinct feature. After the K-means clustering model is applied, we calculate the silhouette score for each topic. This involves using a pre-determined list of $K$ values and computing the mean score across all these values at the end of the cycle, providing a quantitative measure of ESG topic representativeness within the industry. The integration of this data into our framework allows for the analysis of ESG topics specifically within technology companies, with the results visually displayed through a heatmap in Figure \ref{fig:measurementsub1}. This approach systematically quantifies how well different ESG strategies represent the typical practices within a given industry sector.

\subsubsection{Cross-Sector Distinctiveness (y-Axis)}
\begin{figure}[h]
	\centering
	\includegraphics[width=0.6\textwidth]{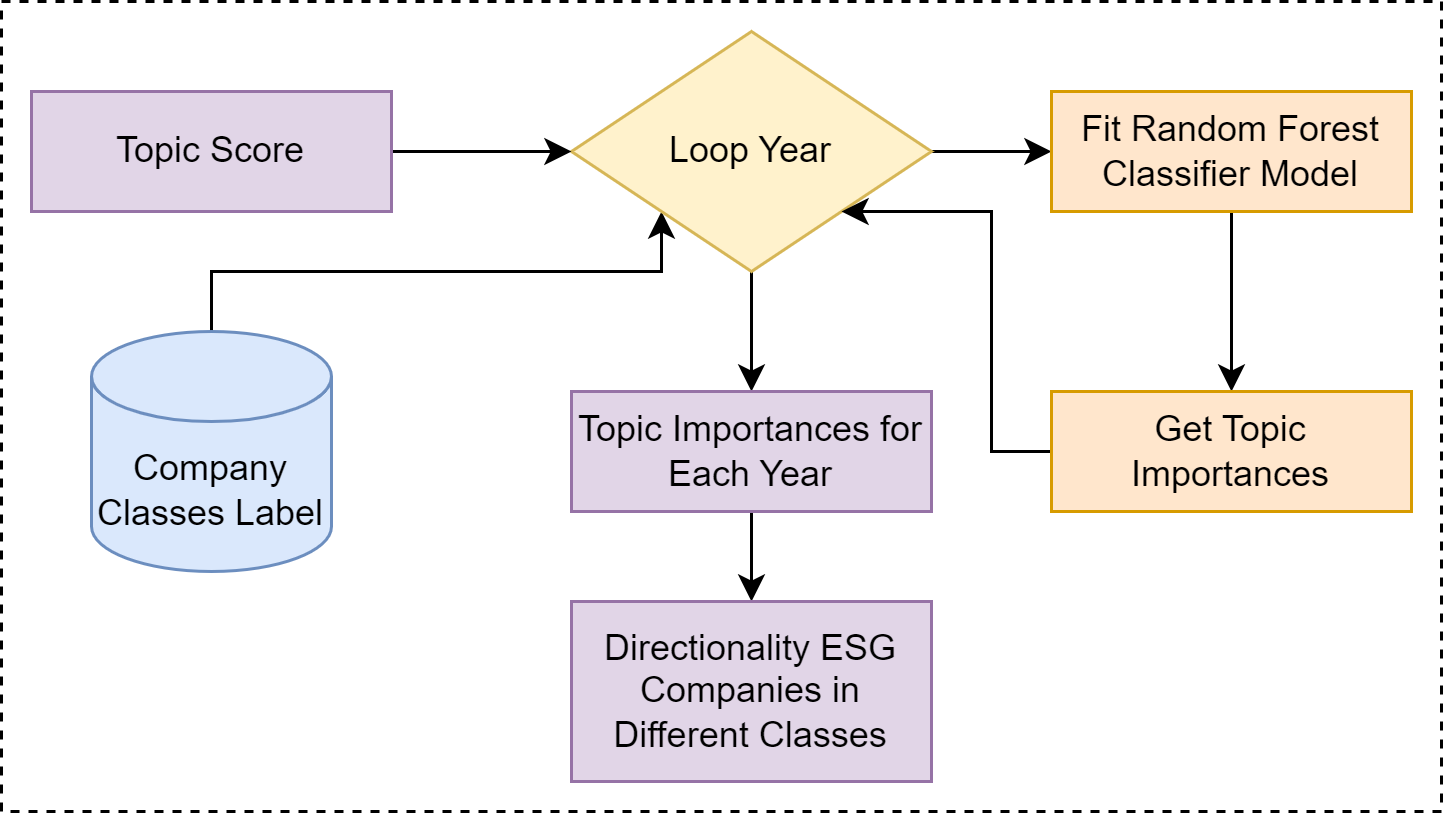}
	\caption{Scrutiny of Directionality for ESG Companies Cross-Sector Distinctiveness}
	\label{fig:diffind}
\end{figure}

To effectively assess ESG performance across different sectors, our framework integrates company class labels such as industry type, service domain, and headquarters country. These labels are crucial for enhancing the analysis of ESG topic importance across various categories. An iterative process is employed annually to train a random forest classifier, utilizing these class labels alongside the pre-compiled ESG data structure. This methodology allows us to derive ESG topic importance that correlates with the designated labels for each year, thereby enhancing the precision of ESG measurements across diverse company classes. The entire process is depicted in Figure~\ref{fig:diffind}. By applying this approach within our framework to analyze ESG topics across different classes of technology service areas, the outcomes are visually presented through a heatmap in Figure \ref{fig:measurementsub2}. This visual representation highlights the distinctiveness of ESG practices across various sectors, illustrating the unique challenges and opportunities in each.

\subsection{ESG Strategy Framework}
Our ESG strategy framework enhances management analysis capabilities by enabling holistic assessments grounded in quantified ESG topics. The framework extends ESG topic measurement across categories, identifying distinctive companies for targeted management research. This methodology also facilitates deep investigations and strategic analysis of key firms in niche markets, uncovering unique ESG practices distinct from other categories.

Our framework can identify representative firms within the industry, enabling comparative studies of ESG performance. This approach illuminates areas of excellence and needed improvements, allowing companies to allocate resources effectively and enhance sustainability practices. By benchmarking against industry standards and identifying flagship firms, companies can emulate best practices and understand the business benefits of robust ESG performance. Additionally, analyzing ESG topic scores and clustering outcomes helps firms recognize emerging trends and adapt strategically to evolving industry and regulatory demands.

\begin{figure}[h]
	\centering
	\includegraphics[width=0.8\textwidth]{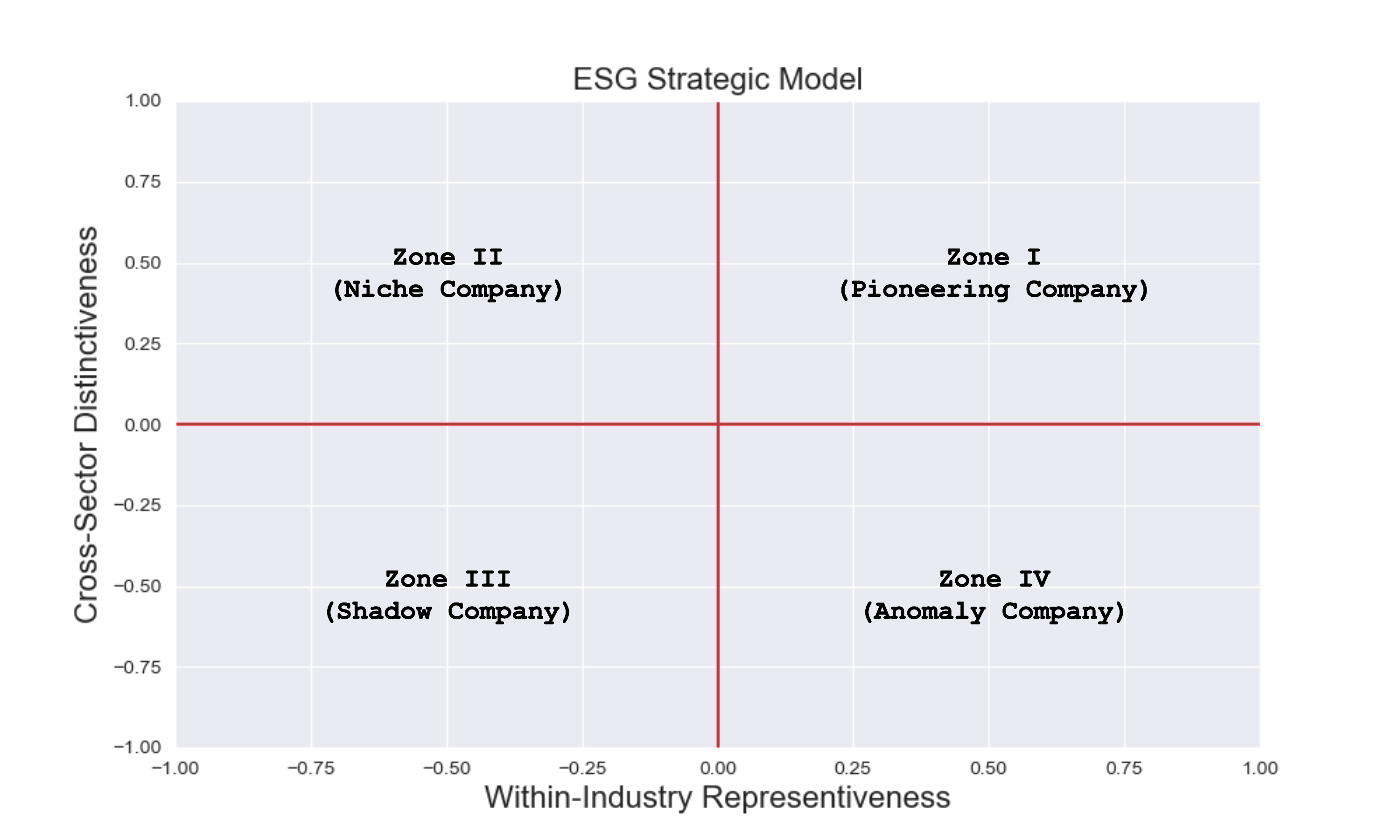}
	\caption{ESG Strategic Model}
	\label{fig:ESGstrmodel}
\end{figure}

The framework also enables analysis of firms' ESG reports across sectors, identifying leaders in each domain and facilitating cross-comparative analysis and benchmarking. This analysis helps industry participants recognize success factors and critical ESG topics, guiding strategic decision-making. Companies can then prioritize and allocate resources to impactful areas, crafting domain specific ESG strategies that address unique challenges and opportunities. Aligning ESG topic importance with sector-specific demands allows organizations to develop targeted sustainability strategies, enhancing ESG outcomes and meeting stakeholder expectations effectively. By integrating the within-class scores from the entire tech industry and the cross-class importance from hardware, software, and service categories, we depict the 2x2 ESG strategic model in Figure~\ref{fig:ESGstrmodel}.

In summary, the ESG Strategy Framework presented in our study offers a novel tool for enhancing corporate sustainability efforts within the industry and across various sectors. By facilitating quantifiable assessments of ESG topics, our framework allows for a deeper understanding of company performance in environmental, social, and governance domains. This enables targeted research, strategic benchmarking, and the identification of best practices, which are instrumental in driving industry-wide improvements and sustainable outcomes. Moreover, the framework's ability to adapt to sector-specific challenges ensures that companies can effectively align their strategies with contemporary demands and stakeholder expectations. By integrating both within-class scores and cross-class importance, the framework not only supports strategic decision-making but also fosters a culture of continuous improvement and innovation within the field of ESG reporting. Thus, it serves as a critical asset for companies aiming to secure a competitive edge and achieve long-term sustainability in an increasingly regulated and normative global market.


\section{Experiment Results and Analysis}\label{sec:exp}
In order to showcase the efficacy of the proposed framework, a comprehensive dataset consisting of ESG reports from 244 technology companies was collected from 2001 to 2022, amounting to a total of 1126 reports sourced from verified official channels. Leveraging the proposed framework, the corresponding ESG topic structure dataset was generated. Notably, this structured ESG topic dataset is the first of its kind effectively organizing the development of ESG topics based on companies' official ESG reports in the 21st century.

\subsection{Identification Results} \label{sec:idenR}

\begin{figure}[]
	\centering

    \begin{subfigure}[b]{0.31\textwidth}
		\includegraphics[width=\textwidth]{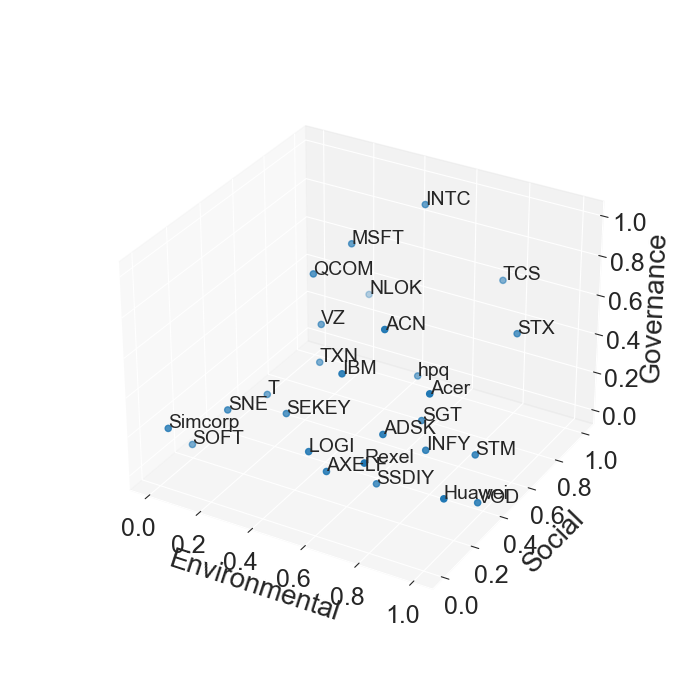}
		\caption{2011 ESG 3-D Plot}
		\label{fig:3dsub2011}
	\end{subfigure}
	\hfill
    \begin{subfigure}[b]{0.31\textwidth}
		\includegraphics[width=\textwidth]{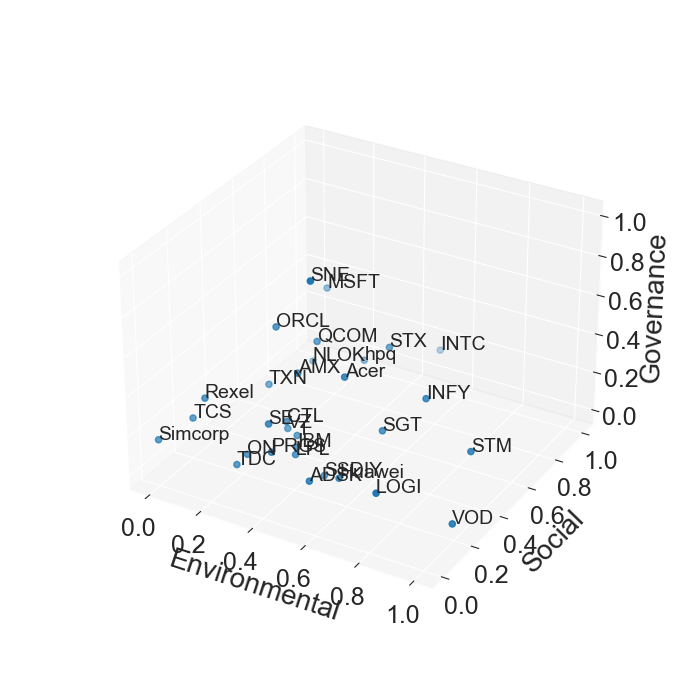}
		\caption{2012 ESG 3-D Plot}
		\label{fig:3dsub2012}
	\end{subfigure}
	\hfill
	\begin{subfigure}[b]{0.31\textwidth}
		\includegraphics[width=\textwidth]{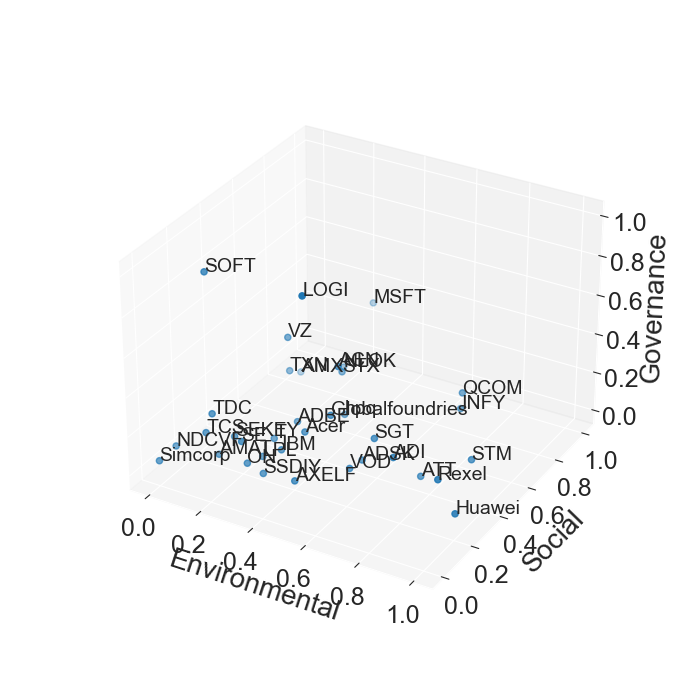}
		\caption{2013 ESG 3-D Plot}
		\label{fig:3dsub2013}
	\end{subfigure}
 
	\begin{subfigure}[b]{0.31\textwidth}
		\includegraphics[width=\textwidth]{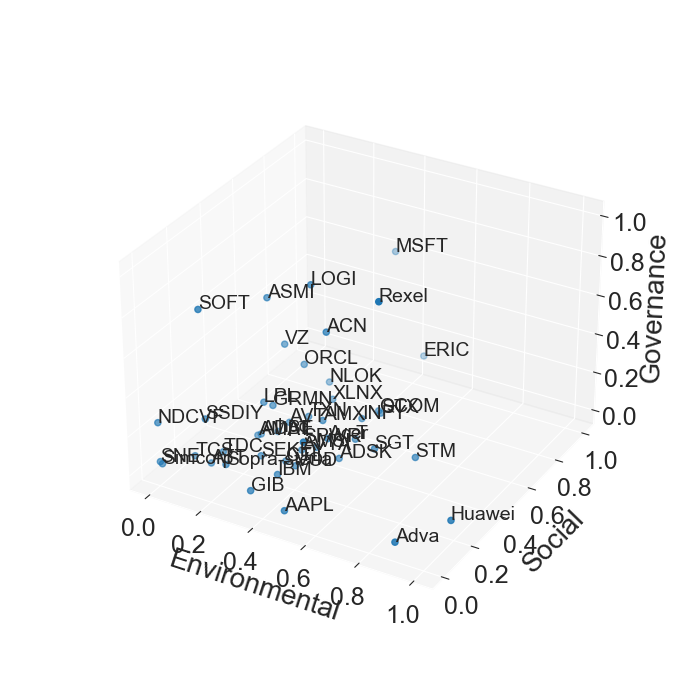}
		\caption{2014 ESG 3-D Plot}
		\label{fig:3dsub2014}
	\end{subfigure}
	\hfill
    \begin{subfigure}[b]{0.31\textwidth}
		\includegraphics[width=\textwidth]{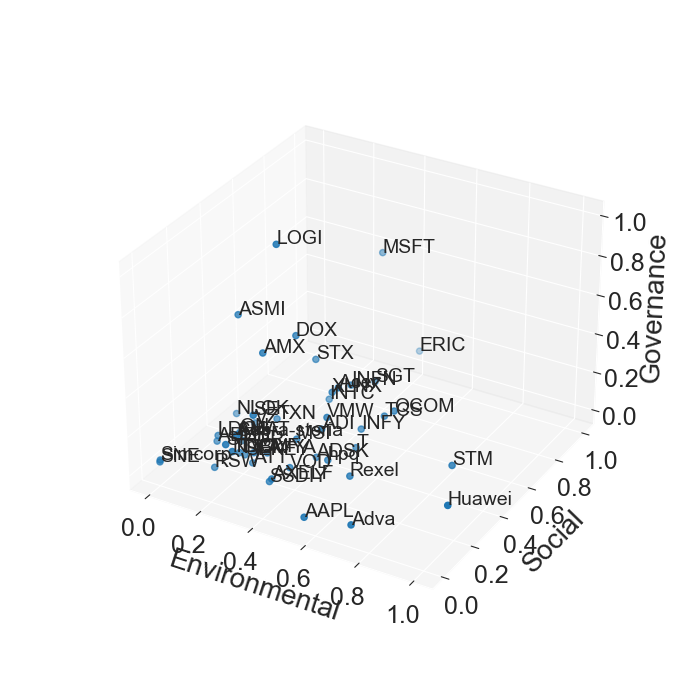}
		\caption{2015 ESG 3-D Plot}
		\label{fig:3dsub2015}
	\end{subfigure}
	\hfill
	\begin{subfigure}[b]{0.31\textwidth}
		\includegraphics[width=\textwidth]{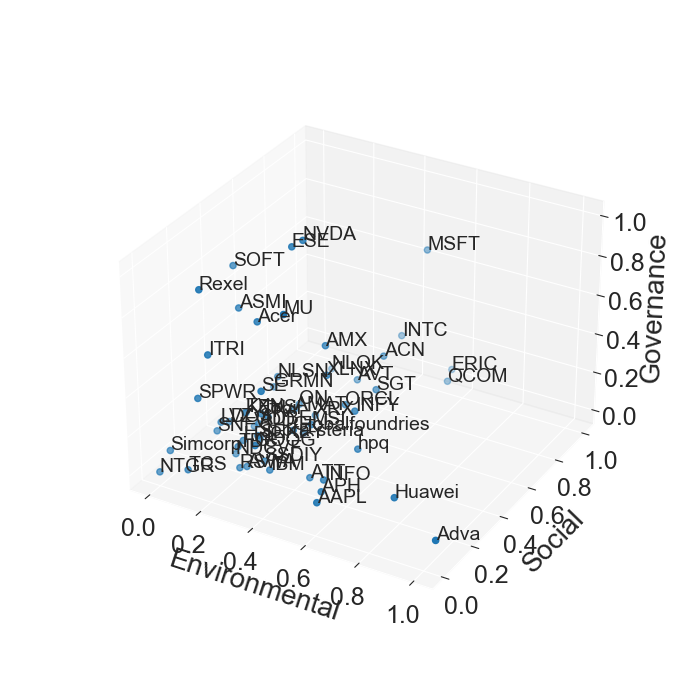}
		\caption{2016 ESG 3-D Plot}
		\label{fig:3dsub2016}
	\end{subfigure}
	
	\begin{subfigure}[b]{0.31\textwidth}
		\includegraphics[width=\textwidth]{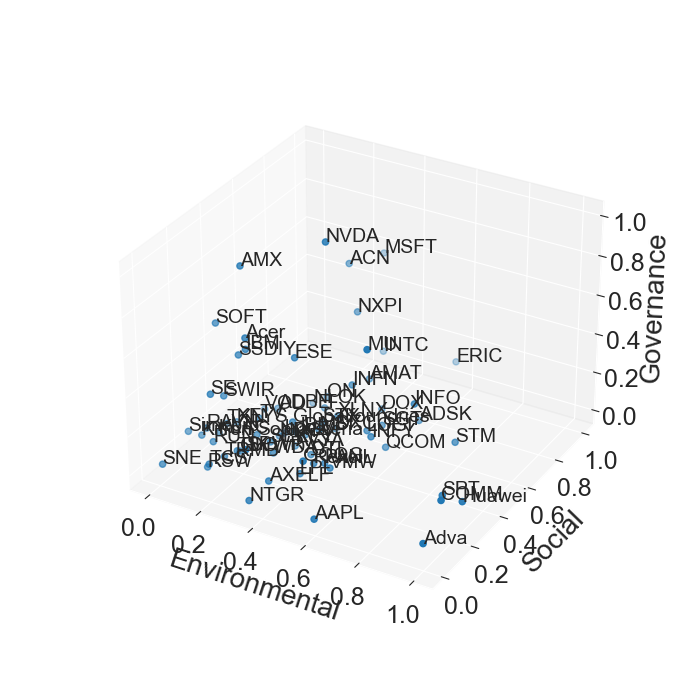}
		\caption{2017 ESG 3-D Plot}
		\label{fig:3dsub2017}
	\end{subfigure}
	\hfill
    \begin{subfigure}[b]{0.31\textwidth}
		\includegraphics[width=\textwidth]{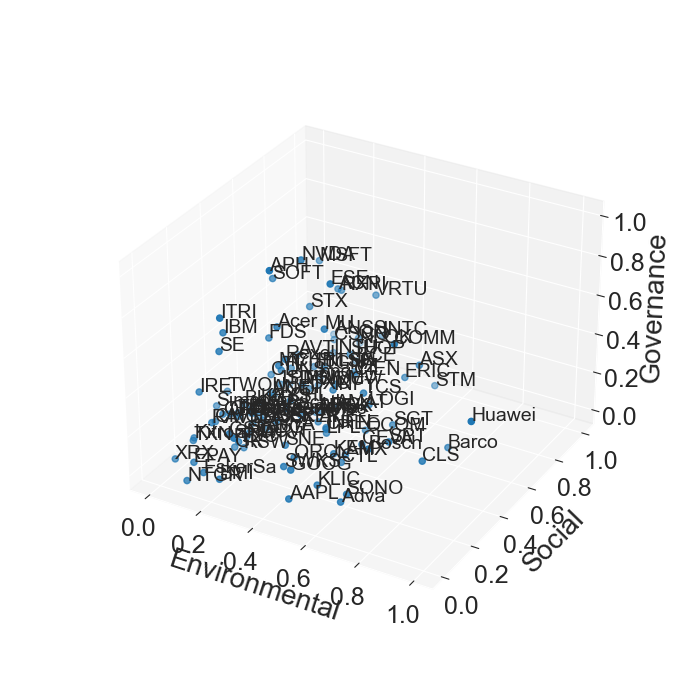}
		\caption{2018 ESG 3-D Plot}
		\label{fig:3dsub2018}
	\end{subfigure}
	\hfill
	\begin{subfigure}[b]{0.31\textwidth}
		\includegraphics[width=\textwidth]{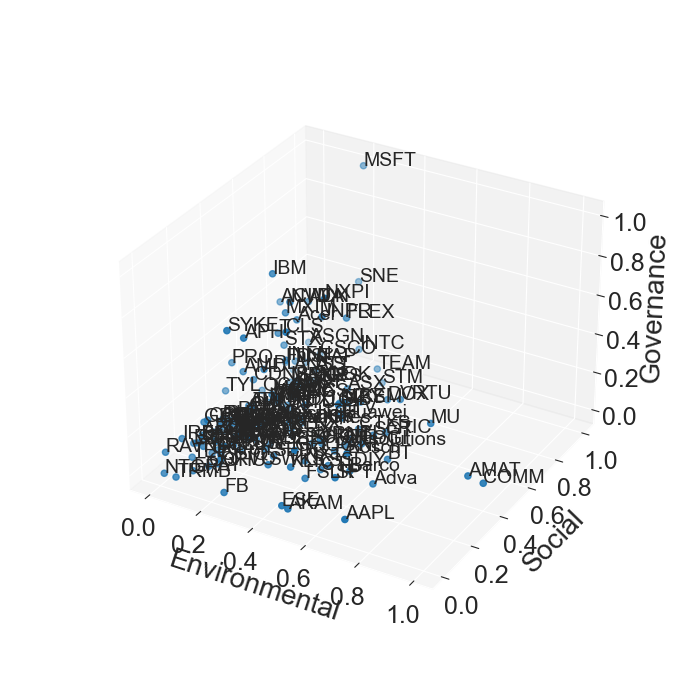}
		\caption{2019 ESG 3-D Plot}
		\label{fig:3dsub2019}
	\end{subfigure}
	
	\begin{subfigure}[b]{0.31\textwidth}
		\includegraphics[width=\textwidth]{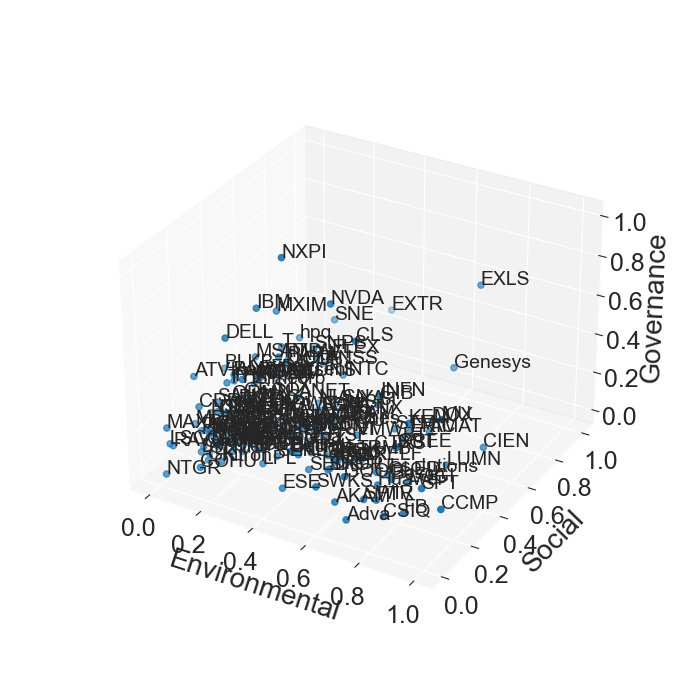}
		\caption{2020 ESG 3-D Plot}
		\label{fig:3dsub2020}
	\end{subfigure}
	\hfill
    \begin{subfigure}[b]{0.31\textwidth}
		\includegraphics[width=\textwidth]{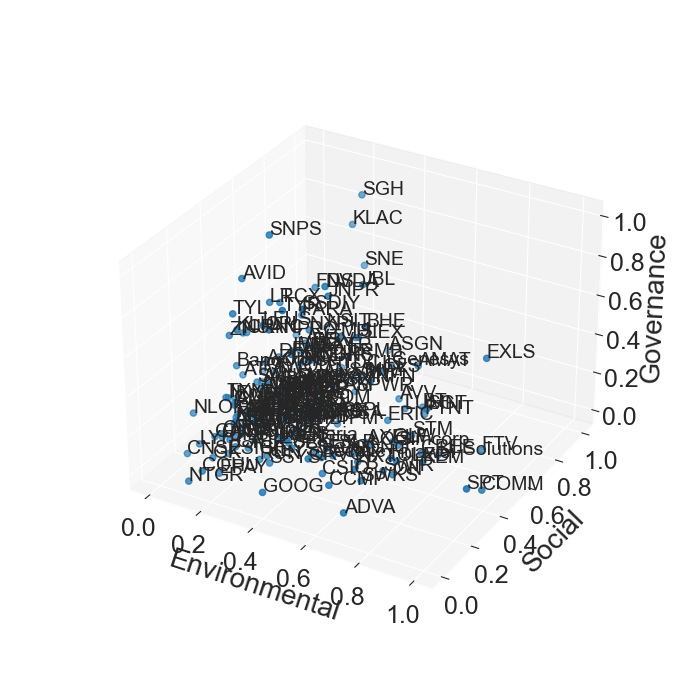}
		\caption{2021 ESG 3-D Plot}
		\label{fig:3dsub2021}
	\end{subfigure}
    \hfill
    \begin{subfigure}[b]{0.31\textwidth}
		\includegraphics[width=\textwidth]{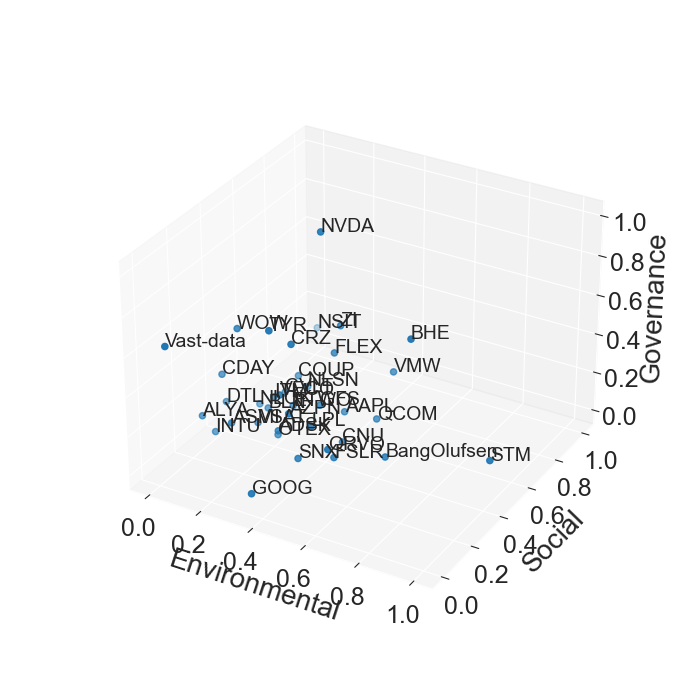}
		\caption{2022 ESG 3-D Plot}
		\label{fig:3dsub2022}
	\end{subfigure}
	
	\caption{2011 to 2022 ESG 3-D Plots}
	\label{fig:ESGplots}
\end{figure}

Once the structured ESG topic scores were obtained an ESG plot was employed to analyze the dynamics of ESG development across the environmental, social, and governance dimensions of companies, as depicted in Figure.~\ref{fig:ESGplots}. After normalizing the data, this visual representation provides insights into the evolving landscape of ESG practices and aids the identification of key trends and patterns over time. By analyzing the three-dimensional ESG evolutionary trends of technology sector firms in each country and region over the period from 2011 to 2022, there is an evolutionary process toward homogeneity in regard to the three dimensions of responsibility: low environmental, medium governance, and high social. The homogenization effect of companies is also visible in ESG reports.

\begin{figure}[]
	\centering
	
	\begin{subfigure}[b]{0.31\textwidth}
		\includegraphics[width=\textwidth]{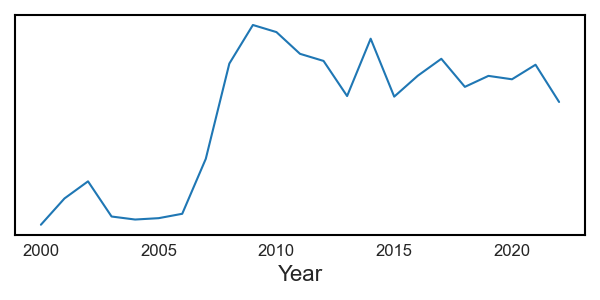}
		\caption{Carbon Emissions}
		\label{fig:sub5}
	\end{subfigure}
	\hfill
	\begin{subfigure}[b]{0.31\textwidth}
		\includegraphics[width=\textwidth]{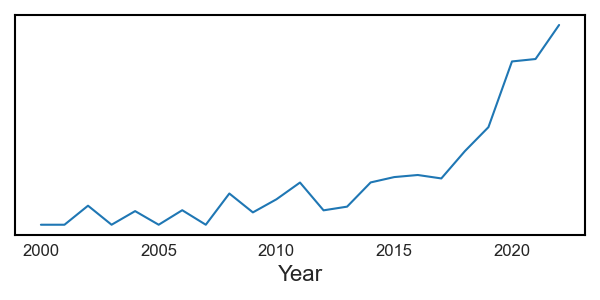}
		\caption{Data Privacy}
		\label{fig:sub1}
	\end{subfigure}
	\hfill
	\begin{subfigure}[b]{0.31\textwidth}
		\includegraphics[width=\textwidth]{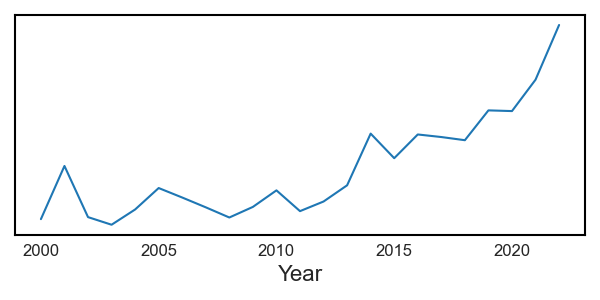}
		\caption{Artificial Intelligence}
		\label{fig:sub3}
	\end{subfigure}

    \begin{subfigure}[b]{0.31\textwidth}
		\includegraphics[width=\textwidth]{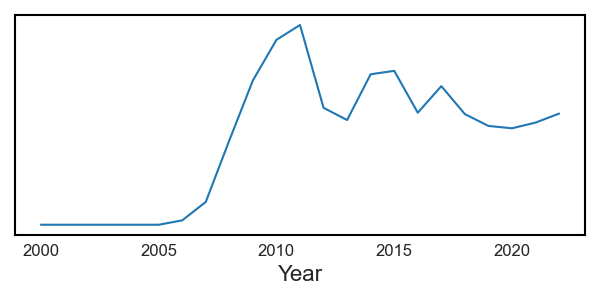}
		\caption{Carbon Footprint}
		\label{fig:sub5}
	\end{subfigure}
	\hfill
	\begin{subfigure}[b]{0.31\textwidth}
		\includegraphics[width=\textwidth]{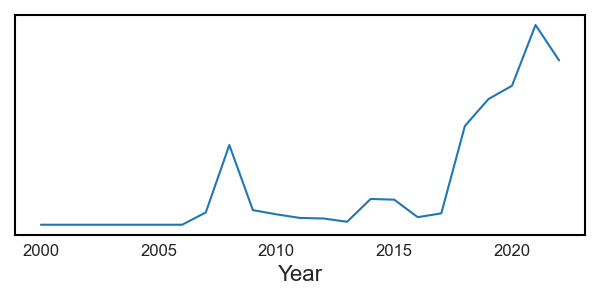}
		\caption{Data Security}
		\label{fig:sub1}
	\end{subfigure}
	\hfill
	\begin{subfigure}[b]{0.31\textwidth}
		\includegraphics[width=\textwidth]{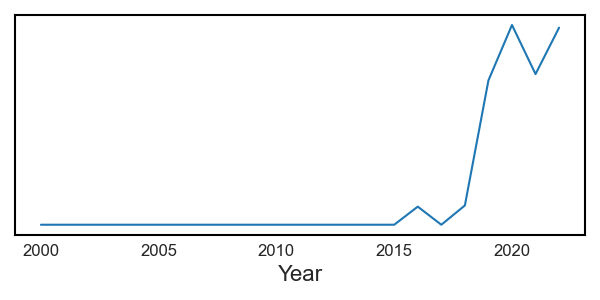}
		\caption{Artificial Intelligence Ethics}
		\label{fig:sub3}
	\end{subfigure}

    \begin{subfigure}[b]{0.31\textwidth}
		\includegraphics[width=\textwidth]{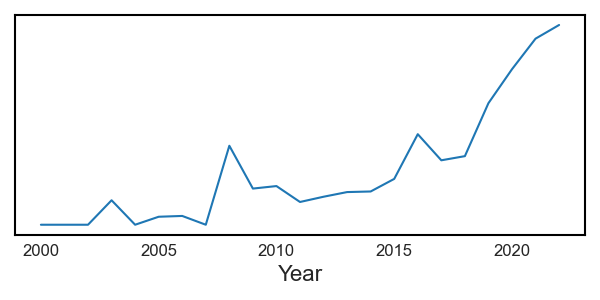}
		\caption{Carbon Neutral}
		\label{fig:sub5}
	\end{subfigure}
	\hfill
	\begin{subfigure}[b]{0.31\textwidth}
		\includegraphics[width=\textwidth]{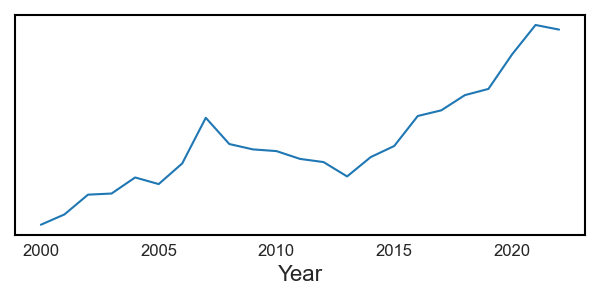}
		\caption{Diversity}
		\label{fig:sub1}
	\end{subfigure}
	\hfill
	\begin{subfigure}[b]{0.31\textwidth}
		\includegraphics[width=\textwidth]{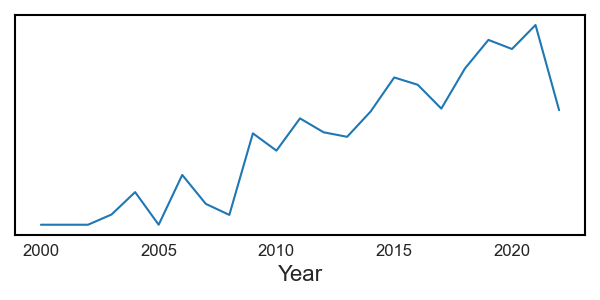}
		\caption{Community Engagement}
		\label{fig:sub3}
	\end{subfigure}

    \begin{subfigure}[b]{0.31\textwidth}
		\includegraphics[width=\textwidth]{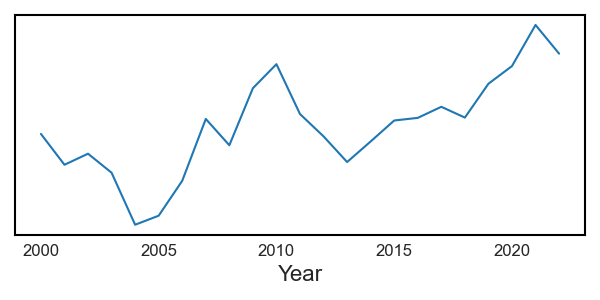}
		\caption{Climate Change}
		\label{fig:sub5}
	\end{subfigure}
	\hfill
	\begin{subfigure}[b]{0.31\textwidth}
		\includegraphics[width=\textwidth]{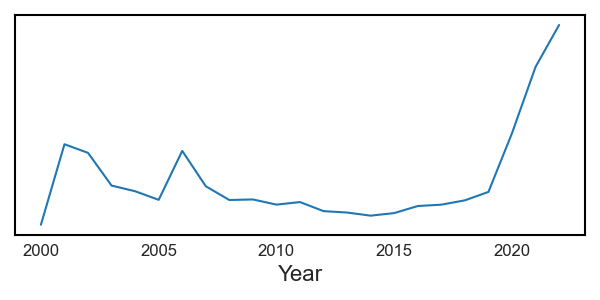}
		\caption{Equity}
		\label{fig:sub1}
	\end{subfigure}
	\hfill
	\begin{subfigure}[b]{0.31\textwidth}
		\includegraphics[width=\textwidth]{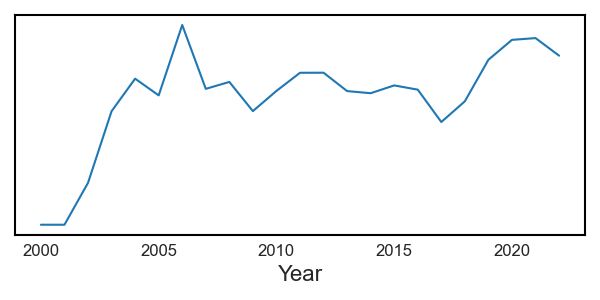}
		\caption{Corporate Governance}
		\label{fig:sub3}
	\end{subfigure}

    \begin{subfigure}[b]{0.31\textwidth}
		\includegraphics[width=\textwidth]{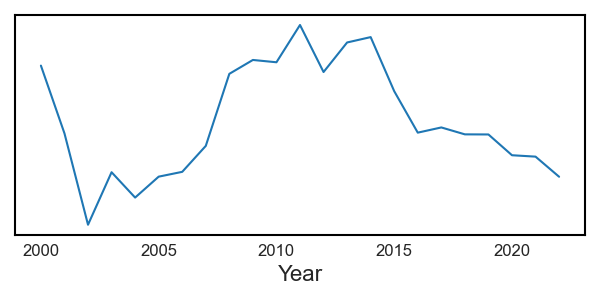}
		\caption{Energy Efficiency}
		\label{fig:sub5}
	\end{subfigure}
	\hfill
	\begin{subfigure}[b]{0.31\textwidth}
		\includegraphics[width=\textwidth]{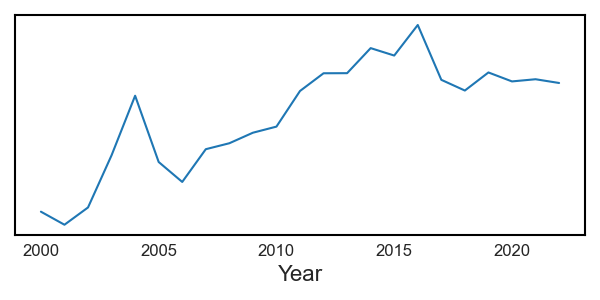}
		\caption{Human Rights}
		\label{fig:sub1}
	\end{subfigure}
	\hfill
	\begin{subfigure}[b]{0.31\textwidth}
		\includegraphics[width=\textwidth]{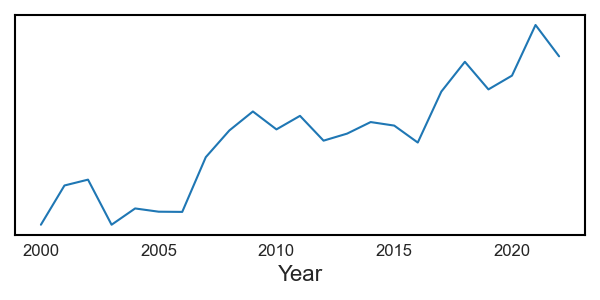}
		\caption{Employee Engagement}
		\label{fig:sub3}
	\end{subfigure}

    \begin{subfigure}[b]{0.31\textwidth}
		\includegraphics[width=\textwidth]{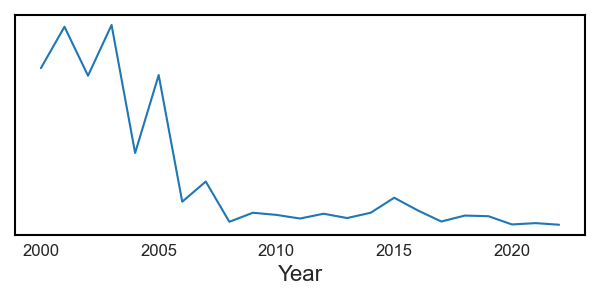}
		\caption{Environmental Management}
		\label{fig:sub5}
	\end{subfigure}
	\hfill
	\begin{subfigure}[b]{0.31\textwidth}
		\includegraphics[width=\textwidth]{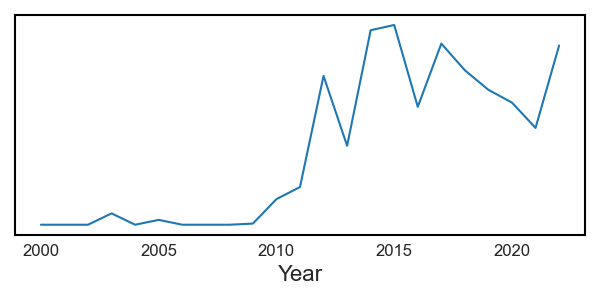}
		\caption{Responsible Sourcing}
		\label{fig:sub1}
	\end{subfigure}
	\hfill
	\begin{subfigure}[b]{0.31\textwidth}
		\includegraphics[width=\textwidth]{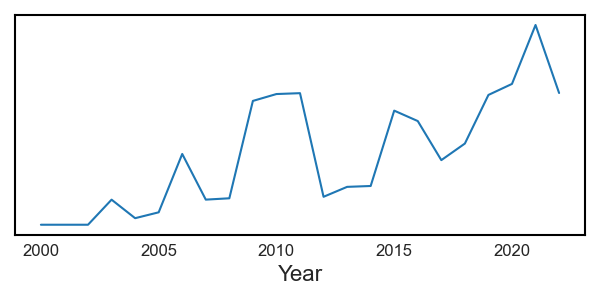}
		\caption{Governance Ethics}
		\label{fig:sub3}
	\end{subfigure}

    \begin{subfigure}[b]{0.31\textwidth}
		\includegraphics[width=\textwidth]{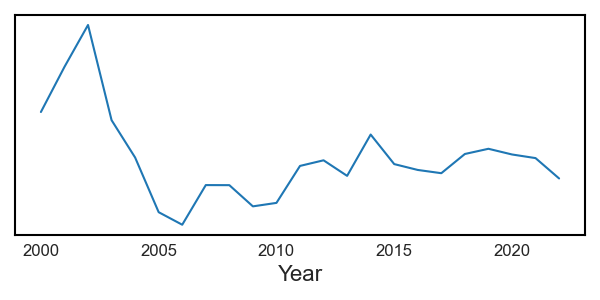}
		\caption{Waste Management}
		\label{fig:sub5}
	\end{subfigure}
	\hfill
	\begin{subfigure}[b]{0.31\textwidth}
		\includegraphics[width=\textwidth]{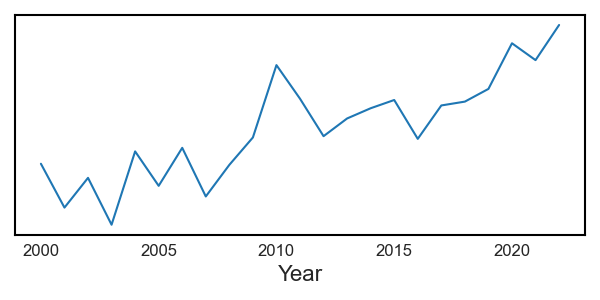}
		\caption{Wellness}
		\label{fig:sub1}
	\end{subfigure}
	\hfill
	\begin{subfigure}[b]{0.31\textwidth}
		\includegraphics[width=\textwidth]{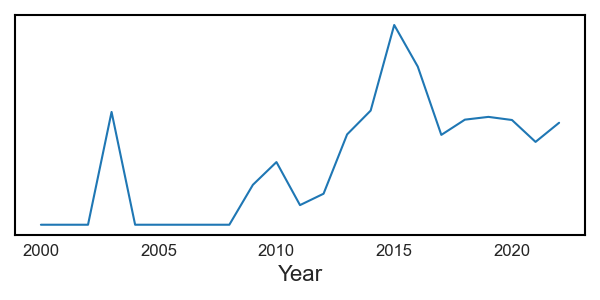}
		\caption{Stakeholder Engagement}
		\label{fig:sub3}
	\end{subfigure}
	
	\caption{ESG Topics Trend Plots, the first column is Environmental, the second column is Social, and the third column is Governance.}
	\label{fig:Topicplots}
\end{figure}

Another series of plots demonstrates ESG development in regard to topic, as shown in Figure.~\ref{fig:Topicplots}. Each ESG topic's change rate is drawn to show their weight change during the whole period from 2001 to 2022. The trend of each topic fits the evolution of low environmental, medium governance, and high social dimensions over time. For instance, the level of importance of the topic of AI ethics and AI research shows exponential growth after 2015. Further, the level of importance of the topic of data privacy and data security shows a high rate of growth after 2015. The level of importance of the topic of responsible sourcing and supplier engagement shows rapid growth after 2010, but not much growth in recent years. Moreover, the level of importance of environmental management and waste management has continued to decline, reaching a relatively stable level after 2005.

\subsection{Measurement Results} \label{sec:measR}

Upon the successful identification of ESG topic data, we apply our proposed measurement methodology to yield quantified outcomes. In the following section, we provide a comprehensive presentation of the three distinct measurement outcomes as discussed in Section~\ref{sec:meas}.

\begin{figure}
	\centering
	
	\begin{subfigure}{0.38\textwidth}
		\includegraphics[width=\textwidth]{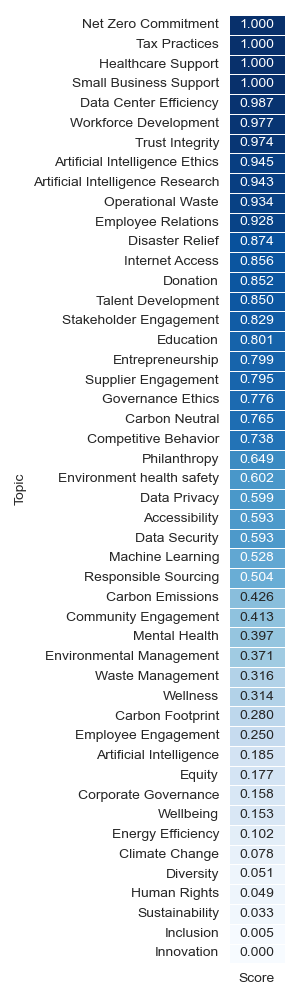}
		\caption{ESG Topic Measurement \\
  Within the Technology Company \\
  Class in 2020}
		\label{fig:measurementsub1}
	\end{subfigure}
	\hfill
	\begin{subfigure}{0.38\textwidth}
		\includegraphics[width=\textwidth]{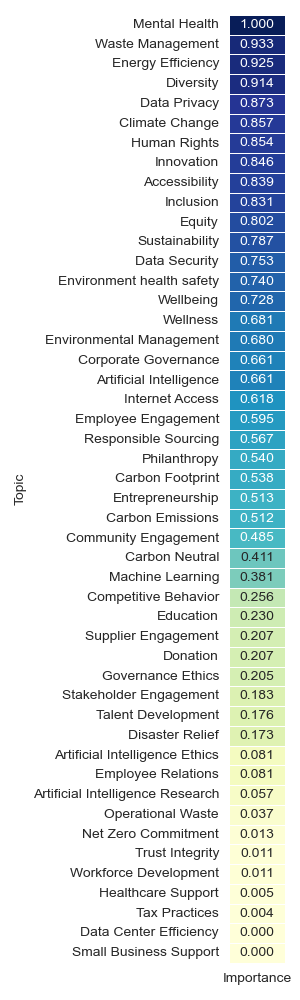}
		\caption{ESG Topic Measurement \\
  Across Technology Company Service \\
  Area Classes in 2020}
		\label{fig:measurementsub2}
	\end{subfigure}

	\caption{ESG Topic Measurement Experimental Results in 2020}
	\label{fig:ESGmeasurementmain}
\end{figure}

It is important to note that all of the ESG reports we amassed are from technology companies. Therefore, our initial measurement involved generating quantile ESG topic scores specific to the technology company class. These scores were subsequently visualized in a heatmap format, and the resultant figures from 2017 to 2020 are exhibited in Figure~\ref{fig:measurementsub1}.

Next, our proposed framework was supplied with the label information corresponding to the service areas of each technology company. We segregated the companies into three classes: hardware companies, software companies, and service companies. These classifications were ascertained based on the companies' primary target areas and their self-identification. Armed with this class label information, we generated ESG topic measurements spanning the different service areas. These measurements are visually represented in Figure~\ref{fig:measurementsub2}.

\subsection{Ranking Results} \label{sec:manaR}

Table~\ref{tab:withinclass} unveils distinct patterns within the technology industry. Business-to-business (B2B) entities such as DXC Technology, Intuit Inc., and CGI Inc. consistently appear at the forefront within the technology company class, underscoring the pivotal role of B2B companies in championing industry sustainability. Notably, Asian tech firms such as LG Display, Samsung SDI, and Huawei also feature in the top five list consistently from 2017 to 2020, accentuating the growing prominence of sustainable practices in Asia. These findings provide critical insights into the B2B sector's role and the rising emphasis on sustainability within the Asian technology industry.

Table~\ref{tab:acrossclass}, illustrating the most directional companies across service areas (software, service, and hardware) from 2017 to 2020, shows intriguing patterns. Microsoft features twice in the software company list, Huawei appears twice in the service company list, and Commscope is listed twice within the hardware company segment. Through a lateral analysis of these companies' ESG reports, we can discern the diverse ESG reporting styles employed by different sectors within the technology industry. Firms exhibit diverse ESG priorities and vary in their level of detail and disclosure in their reports. These differences likely stem from industry-specific challenges and the imperative to meet stakeholder expectations. A scholarly exploration of these discrepancies could further enrich our understanding of the idiosyncrasies of ESG reporting practices across distinct technology industry sectors.

\begin{table}[]
	\centering
	\begin{tabular}{lllll}
		Year        & 2017            & 2018  & 2019   & 2020         \\
		1st Company & DXC             & INTU  & LPL    & GIB          \\
		2nd Company & SOFT            & TGNA  & INTC   & Ul-solutions \\
		3rd Company & Globalfoundries & DXC   & Huawei & ANSS         \\
		4th Company & LOGI            & SSDIY & SSDIY  & CAJ          \\
		5th Company & AMX             & LPL   & hpq    & INTC        
	\end{tabular}
	\caption{ESG Topic Management Within the Technology Company Class from 2017 to 2020}
	\label{tab:withinclass}
\end{table}

\begin{table}[]
	\centering
	\begin{tabular}{lllll}
		Year             & 2017   & 2018   & 2019 & 2020 \\
		Software Company & INFO   & MSFT   & MSFT & BB   \\
		Service Company  & Huawei & Huawei & SNE  & BCE  \\
		Hardware Company & COMM   & COMM   & AMAT & XLNX
	\end{tabular}
	\caption{ESG Topic Management Across Technology Company Service Area Classes from 2017 to 2020}
	\label{tab:acrossclass}
\end{table}


\subsection{Strategic Analysis} \label{sec:straR}

\begin{figure}
	\centering
	
	\begin{subfigure}[b]{0.7\textwidth}
		\includegraphics[width=\textwidth]{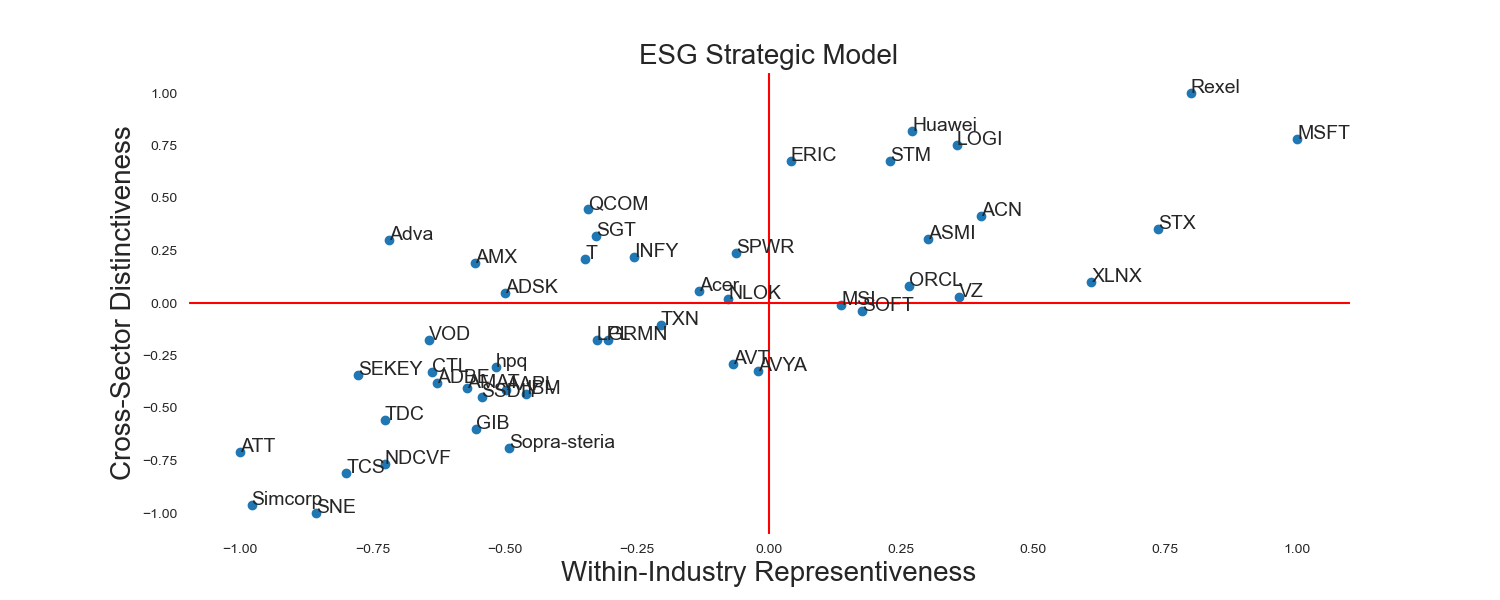}
		\caption{2014 ESG Management Plot}
		\label{fig:sub1}
	\end{subfigure}
	
	\vspace{0.5cm}
	
	\begin{subfigure}[b]{0.7\textwidth}
		\includegraphics[width=\textwidth]{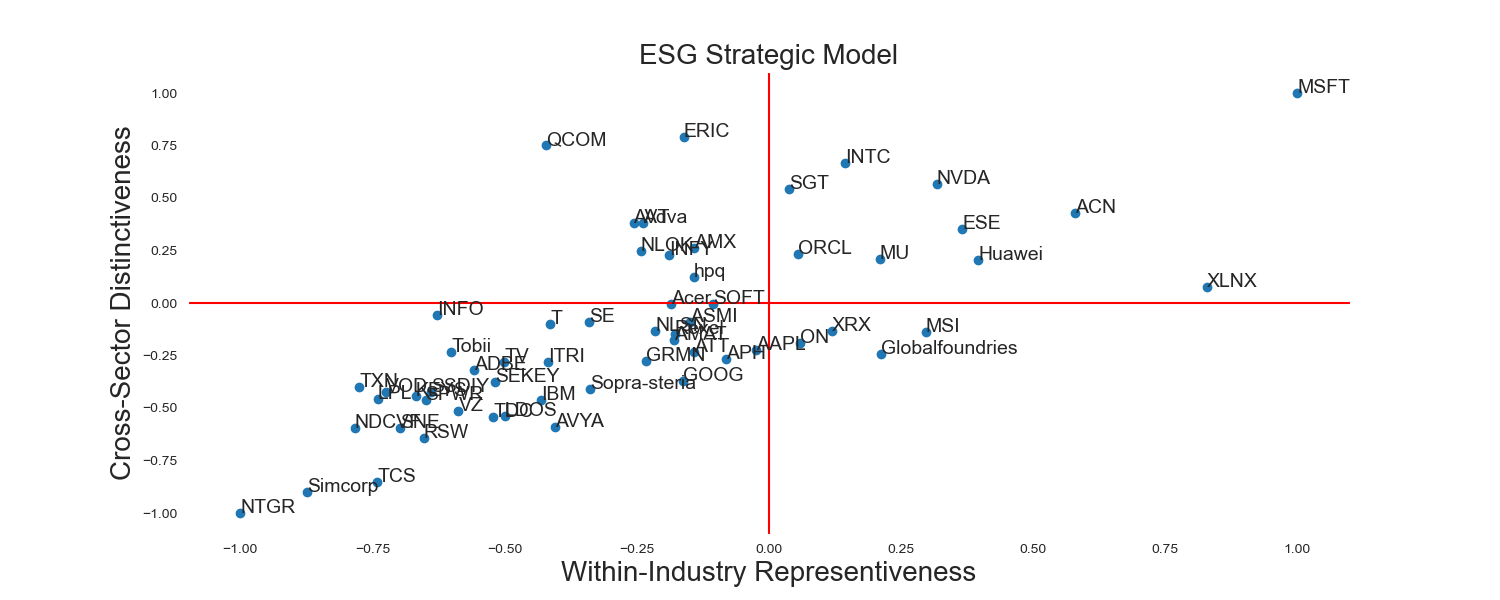}
		\caption{2016 ESG Management Plot}
		\label{fig:sub3}
	\end{subfigure}
	
	\vspace{0.5cm}
	
	\begin{subfigure}[b]{0.7\textwidth}
		\includegraphics[width=\textwidth]{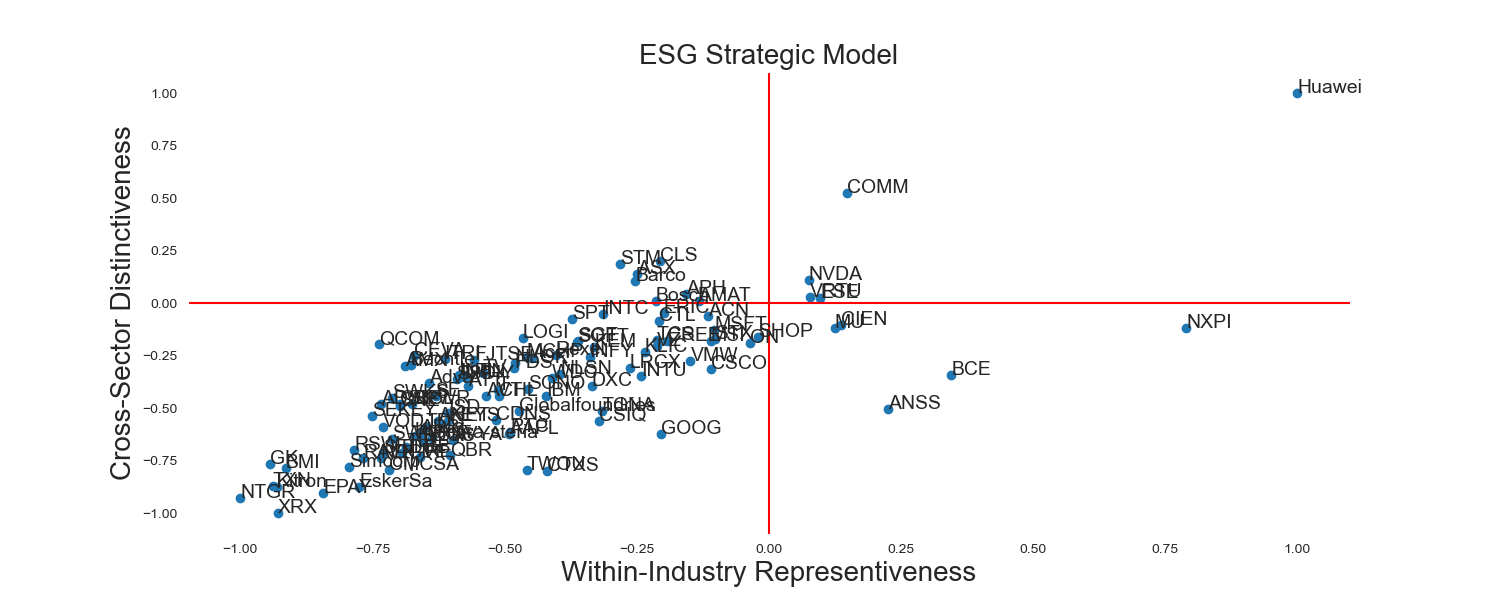}
		\caption{2018 ESG Management Plot}
		\label{fig:sub5}
	\end{subfigure}
	
	\vspace{0.5cm}
	
	\begin{subfigure}[b]{0.7\textwidth}
		\includegraphics[width=\textwidth]{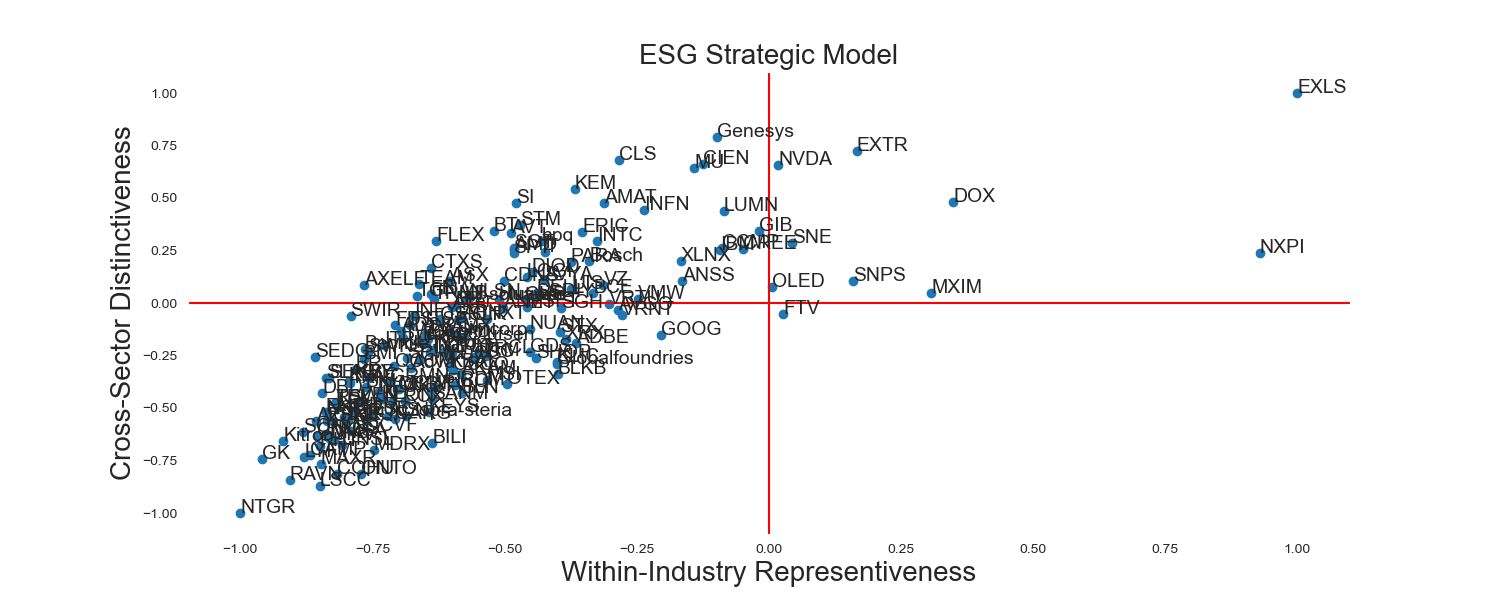}
		\caption{2020 ESG Management Plot}
		\label{fig:sub7}
	\end{subfigure}
	\hfill
	
	\caption{2014, 2016, 2018, and 2020 ESG Strategic Model for Management Plots}
	\label{fig:ESGManagementplots}
\end{figure}

An examination of Figure~\ref{fig:ESGManagementplots}, which delineates the ESG Management Plots from 2014 to 2021, uncovers several notable findings regarding ESG reporting trends within the technology sector.

First, a substantial proportion of firms' ESG reports are found in Zone III, indicative of 'shadow strategy.' Firms within this zone are characterized by their low within-class scores and low cross-class importance, suggesting that they neither command a high level of representativeness within the technology industry nor demonstrate strong distinctiveness across different service areas. Most technology firms appear to strategically align themselves with the dominant trend in ESG reporting, a shadow strategy that allows them to avoid potential misrepresentation, which could lead to adverse implications for ESG investment. This widespread adoption of shadow strategies can be attributed to the mimetic processes, a concept from the theory of institutional isomorphism. In the face of uncertainties or to conform to perceived successful industry standards, technology firms often imitate the ESG practices of their peers, especially those regarded as industry leaders. This mimicry may reduce the burden of developing unique strategies and mitigates risks but leads to a homogenization of ESG practices across the sector, reflecting the broader dynamics of collective rationality that drives firms towards conformity.

Second, the increasing number of reports classified as 'niche strategy' in Zone II from 2014 to 2021 highlights a trend where companies are concentrating on specific ESG topics within their specified domains. These niche companies focus on targeted ESG aspects to improve their sustainability performance in specialized domains. This pattern indicates that more technology sector firms are tailoring their ESG reports to the unique challenges and opportunities of their specific sectors, such as hardware, software, and services in this study. Customizing ESG strategies to sector-specific needs may enhance reputation and efficiency, aligning with stakeholder expectations. As specialization benefits become evident, the trend towards customized ESG reporting in the technology sector is expected to enhance sustainability effectiveness.

A decreasing number of firms appear in Zone I, labeled as 'pioneering strategy' within the technology industry. These firms, with high within-class scores and significant cross-class importance, set within-industry representiveness and cross-sector distinctiveness in ESG reporting. Over time, an increasing number of companies have shifted away from pioneering roles towards Zone II and Zone III, leading to a homogenization of ESG reporting. This trend reflects institutional isomorphism theory, specifically coercive isomorphism, where firms conform to ESG reporting standards due to pressures from dependencies, legal obligations, and societal expectations.

In summary, our analysis highlights diverse ESG reporting strategies within the technology sector. The prevalence of the shadow strategy in Zone III reveals a cautious approach by many firms to mitigate risks associated with mimetic ESG reporting, thereby avoiding potential uncertainty. The rise of niche strategies in Zone II indicates a growing acknowledgment among firms of the necessity to tailor ESG efforts to specific sectoral demands, enhancing both their reputation and operational efficiency. Meanwhile, the diminishing presence of pioneering firms in Zone I reveals that their role in setting high standards and creating unique content for ESG reporting is not becoming a popular trend.

\subsection{Enhanced Insights into ESG Strategic Dimensions}

\begin{table}
\centering
\caption{OLS Regression Results: Relationship between Within-Industry Representativeness and Cross-Sector Distinctiveness}
\label{tab:ols_results}
\begin{tabular}{lcccccc}
\hline
\textbf{Variable} & Coefficient & Std. Error & t-Value & $P>|t|$ & [0.025 & 0.975] \\ \hline
Constant          & -0.2986             & 0.017               & -17.944          & 0.000          & -0.331          & -0.266          \\
Cross-Sector           & 0.6666              & 0.041               & 16.378           & 0.000          & 0.587           & 0.747           \\ \hline
\end{tabular}
\smallskip
\footnotesize{Note: R-squared = 0.461, Adjusted R-squared = 0.459, F-statistic = 268.2, Prob (F-statistic) = 5.20e-44, AIC = 120.4, BIC = 127.9}
\end{table}

In this regression analysis, we study the correlation between Within-Industry Representativeness (x-Axis) and Cross-Sector Distinctiveness (y-Axis) through an OLS regression model. As presented in Table \ref{tab:ols_results}, the model shows a robust R-squared of 0.461, suggesting that about $46.1\%$ of the variance in the dependent variable, within-industry representativeness, can be accounted for by the predictors used.

The constant coefficient is significantly negative (-0.2986), indicating the baseline level of within-industry representativeness when the cross-sector distinctiveness is zero. The coefficient for cross-sector distinctiveness is 0.6666, which is statistically significant with a t-value of 16.378 and a p-value of virtually zero. This strong positive relationship highlights that as the impact sum increases, there is a corresponding increase in the score sum, underscoring a strong positive effect of cross-sector distinctiveness on within-industry representativeness.

The F-statistic of 268.2 with an extremely low probability (5.20e-44) strongly rejects the null hypothesis that the variable cross-sector distinctiveness has no effect, further solidifying the model's reliability. The Durbin-Watson statistic of 1.774 indicates a moderate level of autocorrelation.

These findings suggest a meaningful and statistically significant relationship between the dimensions studied, reinforcing the importance of considering both within-industry representativeness and cross-sector distinctiveness in analyzing ESG strategic frameworks. The results provide a clear quantitative basis for advancing sustainable practices, backed by significant empirical evidence.

\section{Conclusion And Future Work}\label{sec:con}
In this study, we introduce a novel strategic framework designed to derive key insights from corporate ESG reports. Our primary goal was to develop a useful tool that provides assistance in strategic management. We evaluated the utility and effectiveness of this framework by applying it to datasets from technology companies, confirming its strong analytical capabilities and its potential for use in managerial settings.

Our research highlights how the homogenization effect among shadow companies contributes to institutional isomorphism theory in ESG strategy contexts. We found that firms often choose conformity over differentiation through mimetic ESG reporting processes, particularly in the technology industry where many cluster within Zone III to align with established norms. This strategic choice emphasizes the importance of aligning with prevailing standards, even at the potential expense of innovation. Our findings validate institutional isomorphism theory and deepen our understanding of how firms balance topic trends and strategic decisions in ESG reporting within a dynamic business environment.

The use of our model in both academic and industry settings could significantly enhance our knowledge of ESG trends within the industry and across different sectors. Managers or investors can use our framework to perform longitudinal studies, tracking changes and improvements in companies' ESG performance over time. This feature enables decision-makers to monitor the effectiveness of sustainability initiatives and identify areas that require continuous improvement.

Our strategy fuses traditional language processing technology, specifically TF-IDF, culminating in a lightweight, user-friendly model. In light of the fast-paced advancement of large language models based on transformer architecture, future work should explore their integration into our framework. Such a merger could potentially amplify the model's language processing proficiency and reliability, thus boosting our capability to scrutinize ESG development in a way that is not just limited to ESG reports, but extends to all ESG-related online information.

Potential future research could also examine the impact of ESG performance on business outcomes, such as financial performance, market valuation, and stakeholder perception. Establishing a causal relationship between ESG development and company performance would substantiate the case for sustainable business practices. Further, our strategic framework could be deployed in sectors beyond technology, evaluating its cross-sectoral applicability and efficacy. Such a holistic approach would foster a deeper understanding of ESG trends and performance across a plethora of industries.


\bibliographystyle{elsarticle-num} 
\bibliography{main}

\section{Supplementary information}

\begin{figure}[ht]
	\centering
	
	\begin{subfigure}[b]{0.7\textwidth}
		\includegraphics[width=\textwidth]{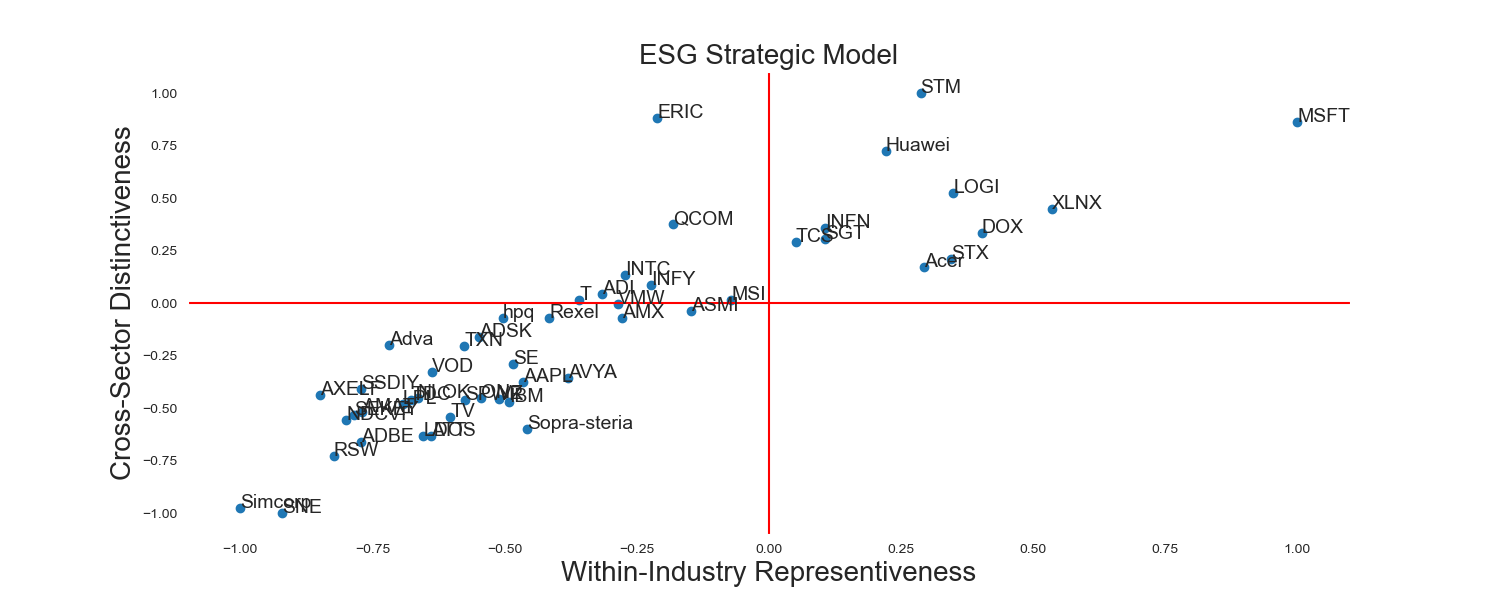}
		\caption{2015 ESG Management Plot}
		\label{fig:sub1}
	\end{subfigure}
	
	\vspace{0.5cm}
	
	\begin{subfigure}[b]{0.7\textwidth}
		\includegraphics[width=\textwidth]{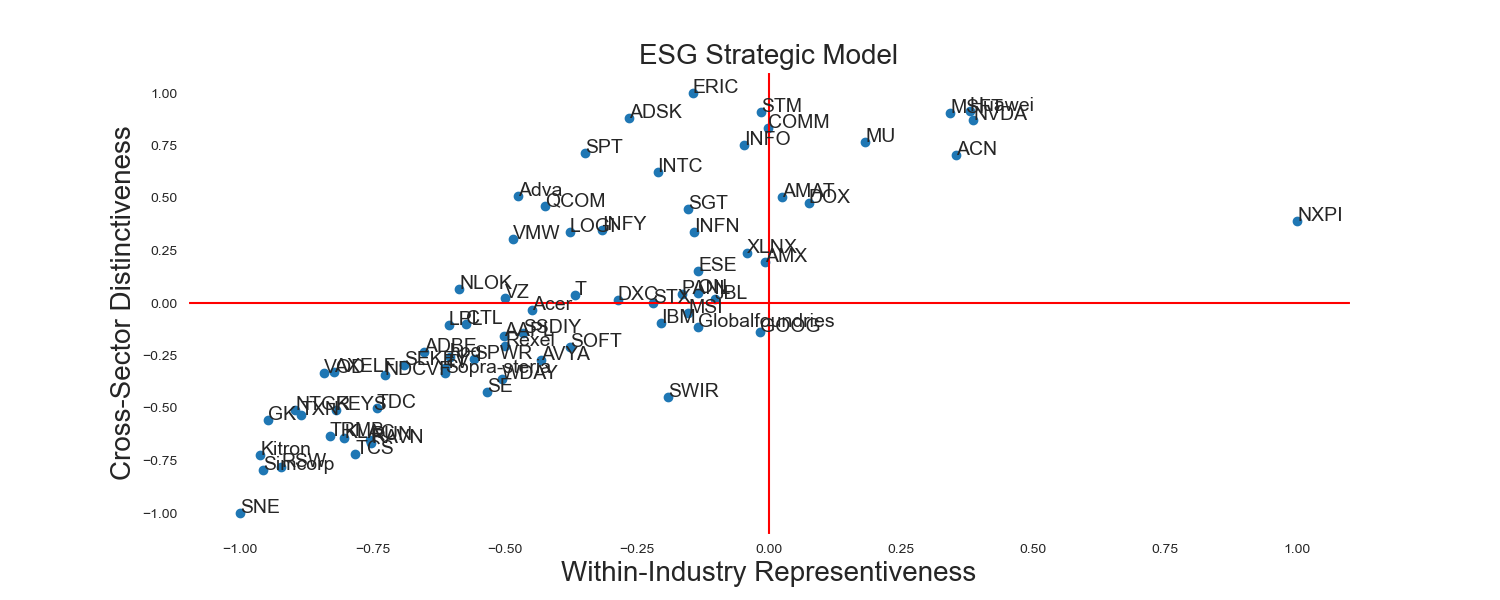}
		\caption{2017 ESG Management Plot}
		\label{fig:sub3}
	\end{subfigure}
	
	\vspace{0.5cm}
	
	\begin{subfigure}[b]{0.7\textwidth}
		\includegraphics[width=\textwidth]{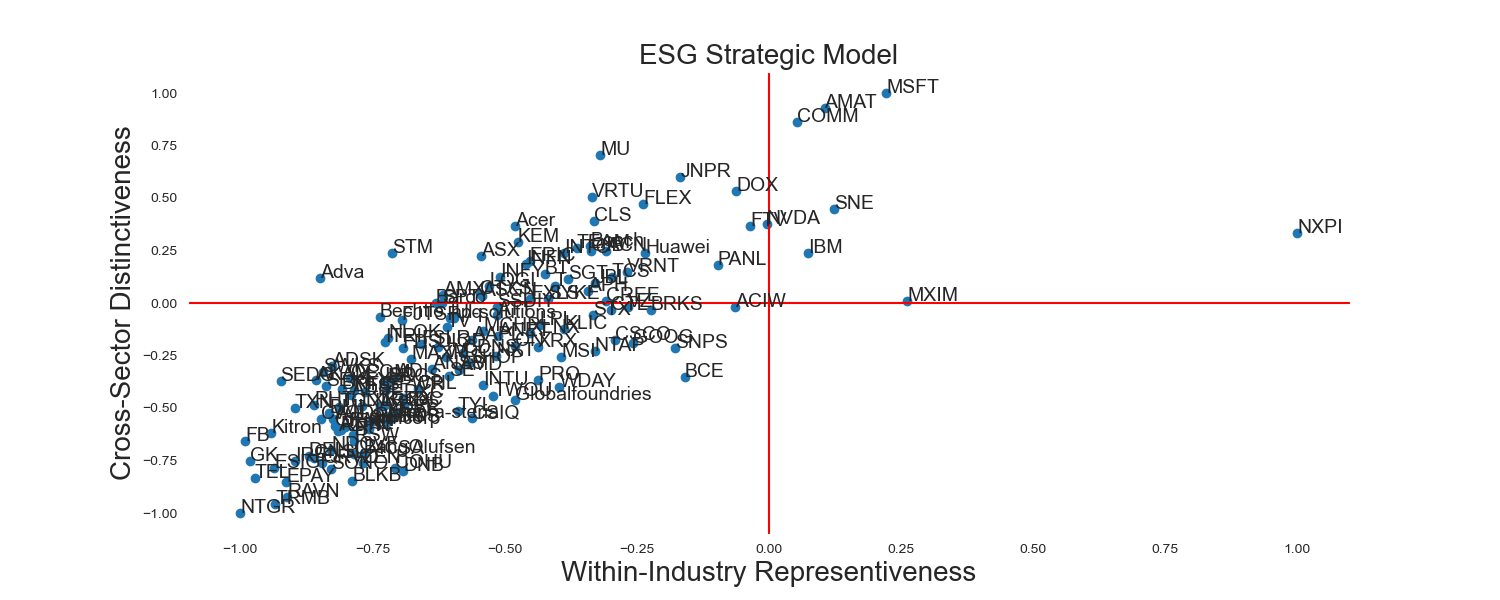}
		\caption{2019 ESG Management Plot}
		\label{fig:sub5}
	\end{subfigure}
	
	\vspace{0.5cm}
	
	\begin{subfigure}[b]{0.7\textwidth}
		\includegraphics[width=\textwidth]{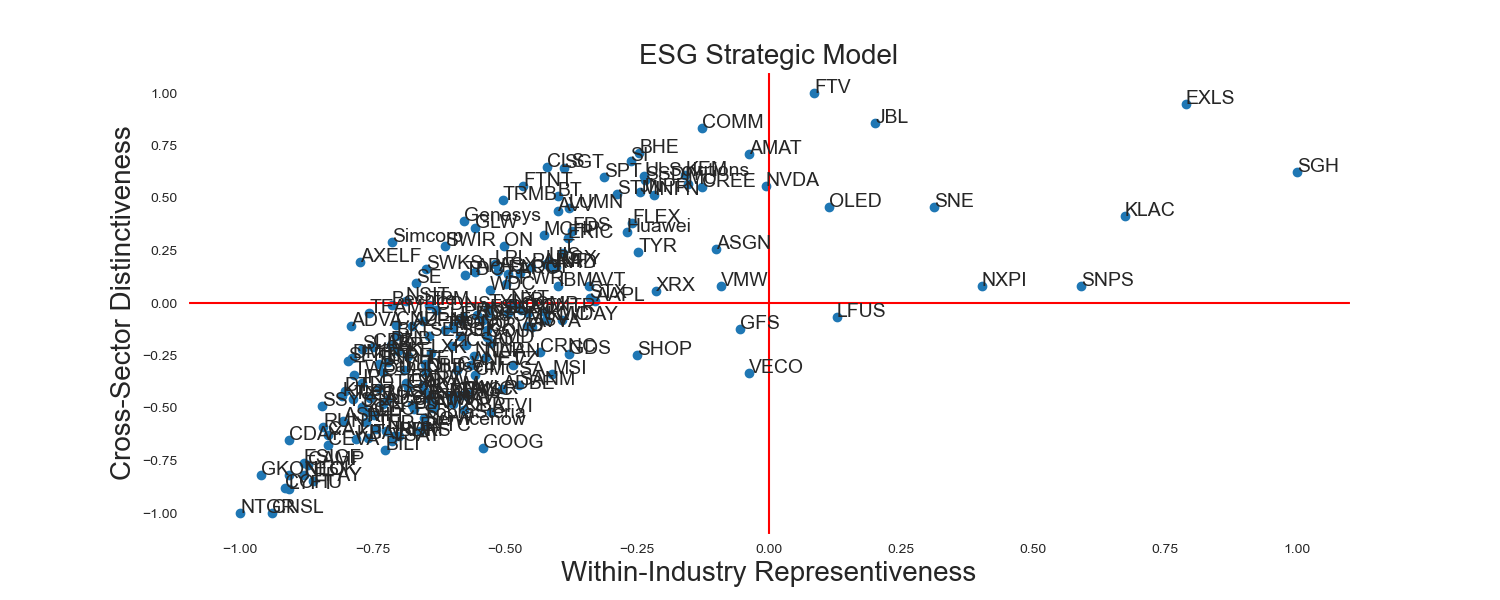}
		\caption{2021 ESG Management Plot}
		\label{fig:sub7}
	\end{subfigure}
	\hfill
	
	\caption{2015, 2017, 2019, and 2021 ESG Strategic Model for Management Plots}
	\label{fig:ESGManagementplots_2}
\end{figure}

\end{document}